\documentclass[preprint2]{aastex}

\slugcomment{Published in ApJ, 766, 116, 2013 (April 10);\\  Erratum: to be published in ApJ, 777, 80, 2013 (Nov 1)}

\shorttitle{Variability in PPNs: II. IRAS 22272+5435 \& 22223+4327}

\shortauthors{Hrivnak et al.}

\begin{document}


\title{STUDIES OF VARIABILITY IN PROTO-PLANETARY NEBULAE: II. LIGHT AND VELOCITY CURVE ANALYSES OF IRAS 22272+5435 and 22223+4327}


\author{Bruce J. Hrivnak\altaffilmark{1, 2}, Wenxian Lu\altaffilmark{1}, Julius Sperauskas\altaffilmark{3}, Hans Van Winckel\altaffilmark{4}, David Bohlender\altaffilmark{5}, and Laimons Za\v{c}s\altaffilmark{3, 6}}

\altaffiltext{1}{Department of Physics and Astronomy, Valparaiso University, 
Valparaiso, IN 46383; bruce.hrivnak@valpo.edu, wen.lu@valpo.edu}
\altaffiltext{2}{Guest investigator, Dominion Astrophysical Observatory, 
Herzberg Institute of Astrophysics, National Research Council of Canada} 
\altaffiltext{3}{Vilnius University Observatory, Ciurlionio 29 Vilnius 2009, Lithuania; julius.sperauskas@ff.vu.lt}
\altaffiltext{4}{Instituut voor Sterrenkunde, K.U. Leuven University, 3001
Leuven (Heverlee) Belgium; Hans.VanWinckel@ster.kuleuven.be}
\altaffiltext{5}{Dominion Astrophysical
Observatory, Herzberg Institute of Astrophysics, National Research
Council of Canada, 5071 West Saanich Road, Victoria, BC V9E 2E7, Canada; David.Bohlender@nrc-cnrc.gc.ca}
\altaffiltext{6}{Faculty of Physics and Mathematics, University of Latvia, Raina bulvaris 19, LV-1586 R\-{i}ga, Latvia; zacs@latnet.lv}

\begin{abstract}

We have carried out a detailed observational study of the light, color, and velocity variations of two bright, carbon-rich proto-planetary nebulae, IRAS 22223+4327 and 22272+5435.  The light curves are based upon our observations from 1994 to 2011, together with published data by Arkhipova and collaborators.  They each display four significant periods, with primary periods for IRAS 22223+4327 and 22272+5435 being 90 and 132 days, respectively.  For each of them, the ratio of secondary to primary period is 0.95, a value much different from that found in Cepheids, but which may be characteristic of post-AGB stars.
Fewer significant periods are found in the smaller radial velocity data sets, but they agree with those of the light curves.
The color curves generally mimic the light curves, with the objects reddest when faintest.
A comparison in seasons when there exist contemporaneous light, color, and velocity curves reveals that the light and color curves are in phase, while the radial velocity curves are $\sim$0.25 {\it P} out of phase with the light curves.  Thus they differ from what is seen in Cepheids, in which the radial velocity curve is 0.50 {\it P} out of phase with the light curve.
Comparison of the observed periods and amplitudes with those of post-AGB pulsation models shows poor agreement, especially for the periods, which are much longer than predicted.
These observational data, particularly the contemporaneous light, color, and velocity curves, 
provide an excellent benchmark for new pulsation models of cool stars in the post-AGB, proto-planetary nebula phase.

\end{abstract}




\section{INTRODUCTION}
\label{intro}

Proto-planetary nebulae (PPNs) represent a transient stage in the evolution of 
intermediate-mass stars between the asymptotic giant branch (AGB) and the
planetary nebula (PN) phases.
The identification and study of these objects began in earnest following the
successful mission of the {\it Infrared Astronomy Satellite} ({\it IRAS}) in 1983.
Candidate PPNs were selected on the basis of their mid-infrared excesses and 
follow-up studies have established the PPN nature of a number of these.
Their variability has been known since early studies of these objects \citep{hri97}.

Light curve studies of individual PPNs have been carried out by Arkhipova 
and collaborators \citep[e.g.,][]{ark00} and by us.  
Two of us recently published a long-term (14 yr) light curve study of 12 carbon-rich
PPNs of F$-$G spectral types \citep[Paper I]{hri10}.  We determined periods for all of them,
ranging from $\sim$40$-$150 days, 
and found inverse relationships between the pulsation period and the effective temperature 
and between the pulsation period and the maximum seasonal variation in brightness.

In this paper, we carry out a more detailed study of two of these carbon-rich PPNs,
IRAS 22223+4327 and 22272+5435, based on extensive light and velocity curve data.
In addition to our published light curve data, we include our newer data and data published 
by others.
Similarly, our new radial velocity data are combined with our published data \citep{zacs09}.
These photometric and velocity data are used to carry out a detailed
pulsational study of these objects.
From these data, we find and compare the periods for the light and velocity variations and
also determine the presence of multiple pulsation periods in the data sets.  
A consistent phase relationship is seen between the light, color, and velocity curves.
Attempts are made to use these to determine the radii and luminosities, but these are not successful.
Finally, comparisons are made with the predictions of non-linear, radial pulsation models.

\section{PROGRAM OBJECTS}
\label{}

Some basic properties of the two program objects are listed in Table~\ref{object_list}. 
The initial evidence for the PPN nature of IRAS 22272+5435 (HD~235858, BD$+54$~2787, SAO~34504, V354 Lac) was presented by \citet{hri91}, whose observations showed a double-peaked spectral energy distribution and a G supergiant spectral type, and who discovered the presence of C$_2$ and C$_3$ in absorption in the optical spectrum.    
Similar properties were later found for IRAS 22223+4327 \citep[V448 Lac, DO~41288;][]{hri95}.

\placetable{object_list}

 As noted above, both are carbon-rich and they both have enhanced {\it s}-process elements due to nucleosynthesis on the
asymptotic giant branch \citep{zacs95,vanwin00,red02}.  
As is typical with PPNs, they each show a double-peaked spectral energy distribution, with one peak in the 
visible/near-infrared arising from the reddened photosphere and a second one in the mid-infrared arising from cool (T$\sim$150 K) dust.  
They are among the small number of objects ($\sim$18 in the Milky Way Galaxy) found to display the unidentified ``21 $\mu$m emission feature'' \citep{volk99};
they also display the ``30 $\mu$m emission feature'' \citep{volk02} and a variety of infrared aromatic features arising in their
dusty circumstellar envelope \citep{hri08}.
The molecular component of their circumstellar envelope is seen  in millimeter-line CO and HCN emission \citep{omo93} and also in near-infrared C$_2$ and CN absorption \citep{bak97}.  These indicate an envelope expansion of 10$-$15 km s$^{-1}$.
The nebulae appear small in the sky as imaged with the {\it Hubble Space Telescope}.  Both show a bright star surrounded by a faint reflection nebula, with sizes of 3.4$\arcsec$$\times$2.1$\arcsec$ \citep{sio08} and 3.5$\arcsec$$\times$3.5$\arcsec$  \citep{uet00} for IRAS 22223+4327 and 22272+5435, respectively.

\section{NEW PHOTOMETRIC OBSERVATIONS}
\label{photo}

New differential photometric observations were carried out at the Valparaiso University Observatory (VUO) during the four seasons from 2008$-$2009 to 2011$-$2012.  (Since our observations in each season are made between May and February, we will hereafter denote them by the earlier of the double years, in which almost all of the observations were made.)  These complement the earlier observations carried out at the VUO from 1994 through 2007.  The newer data were obtained with a new, larger-format detector, an SBIG 6303 CCD with 2048$\times$3072 pixels, which was binned 2$\times$2, with a resulting binned pixel size of 0.70$\arcsec$ on the sky.  
The data were reduced using IRAF\footnote{IRAF is distributed by the National Optical
Astronomical Observatory, operated by the Association for
Universities for Research in Astronomy, Inc., under contract with
the National Science Foundation.}, and aperture photometry was carried out with an aperture of 11$\arcsec$, similar to what was used in Paper I.
Three comparison stars  (C$_1$, C$_2$, C$_3$) were used for each object, with one designated as the primary (C$_1$) and used for the differential photometry measurements.
For IRAS 22223+4327, we used the same comparison stars as listed in Paper I.  
For IRAS 22272+5435, we replaced two of them with brighter stars available with the larger CCD, and we used a new primary comparison star because the previous one appeared to be increasing slightly in brightness by $\sim$0.010 mag over the four recent years of observation.
Standard {\it B}, {\it V}, {\it R$_C$} filters were used, with {\it R$_C$} on the Cousins system. 
Standardized photometry of the program and comparison stars are listed in Table~\ref{phot_std}
and standardized differential magnitudes are listed in Tables~\ref{new_phot_22223} and \ref{new_phot_22272}.
The numbers of new observations are as follows:
for IRAS 22223+4327 -- 166 ({\it B}), 173 ({\it V}), 176 ({\it R$_C$}), and
for IRAS 22272+5435 -- 167 ({\it B}), 170 ({\it V}), 170 ({\it R$_C$}).
The precision of the data is good, with statistical uncertainties for IRAS 22223+4327 of 
$\sigma$ $\le$ 0.005 mag and for IRAS 22272+5435 of $\sigma$ $\le$ 0.010 mag for 
{\it B} and {\it V} and $\sigma$ $\le$ 0.014 mag for {\it R$_C$}.
The lower precision for IRAS 22272+5435 is due to the larger difference in magnitude between it and the comparison star.

\placetable{phot_std}

\placetable{new_phot_22223} 
\placetable{new_phot_22272}

\section{RADIAL VELOCITY OBSERVATIONS AND REDUCTIONS}
\label{velocities}

New radial velocity observations were made from three different observatories in a concerted effort to obtain good time coverage.  These are described below, along with some details of the different data sets.

The radial velocity observations for this program were initially made at the Dominion
Astrophysical Observatory (DAO) with the 1.2-m telescope used at 
the Coud\'{e} focus. 
Observations were made in two observing intervals:
1991 to 1995 (especially 1991 and 1992), and 2007 to 2011.  
The 1991$-$1995 DAO data were obtained with the radial-velocity spectrometer \citep[RVS,][]{fle82}, and they are hereafter referred to as DAO-RVS data.
This instrument used a physical spectral mask of an F star, based upon Procyon, and included
about 340 of the sharpest stellar lines in the wavelength interval from 4000 to
4600 $\AA$. 
The DAO radial velocity measurements from 2007$-$2011 were derived from 
high-resolution CCD spectra in the wavelength interval from 4350 to 4500 $\AA$ 
by cross-correlation with 
high signal-to-noise spectra of IAU radial velocity standards taken with the same instrument. 
They are hereafter referred to as DAO-CCD data. 
Care was taken to insure that the two sets of measurements are on the same velocity system.
The zero-point of the much more numerous RVS observations of standard stars has been adjusted to match that of the DAO photographic data \citep{sca90,sca10}.
The velocities of the CCD velocity standards were taken to be those derived at the DAO photographically and with the RVS. 
These two PPNs, with G spectral types, possess numerous absorption lines that are moderately sharp, 
and this results in an observational precision of $\sim$0.7 km~s$^{-1}$ for the first interval 
and $\sim$0.5 km~s$^{-1}$ for the second.

Radial velocity observations of IRAS 22272+5435 were initiated by Za\v{c}s and Sperauskas in 2005 using the CORAVEL spectrometer of Vilnius University \citep{upg02},  and the observations from 2005 to 2007 have been published \citep{zacs09}.  
Additional observations of this object have continued and observations of IRAS 22223+4327 were initiated in 2008 with the CORAVEL spectrometer mounted on the 1.65 m telescope at the Moletai Observatory (Lithuania). 
Radial velocities were measured using the physical mask containing 1650 slits spread over the spectral interval 3850$-$6400 $\AA$ and based mainly on the Solar spectrum.  Some slits  are centered on the lines  from Arcturus and  Procyon spectra also.
The precision in the data is $\sim$0.6$-$0.7 km~s$^{-1}$. 

Additional radial velocity observations for both objects were also made from 2009 to 2011 with the 
fiber-fed echelle spectrograph Hermes on the 1.2-m Flemish Mercator telescope at La Palma  \citep{rask11}.  
The spectra cover the wavelength range 3900 to 9000 
$\AA$ with a spectral resolution of
$\Delta\lambda$/$\lambda$ $\sim$ 85,000. The data were reduced using a specifically-developed instrument pipeline that made use of 
discrete cross correlation; the template was based on a G star and constructed from the line centers and relative depths of the spectral lines.  
These resultant velocities have a precision of $\sim$0.2 km~s$^{-1}$.
The velocities are scaled to the IAU system through observations of IAU radial velocity standards \citep{udry99}.  

For IRAS 22223+4327, we have obtained a total of 178 observations, 
34 from the first DAO interval (DAO-RVS), 42 from the second (DAO-CCD), 
66 from CORAVEL, and 36 from Hermes.  
These are listed in Table~\ref{tab_22223RV}.
For IRAS 22272+5435, we have obtained a total of 202 unpublished observations, 
33 from the first DAO interval (DAO-RVS), 44 from the second (DAO-CCD), 
79 from CORAVEL, 41 from Hermes, and an additional five observations kindly made for us by R. McClure with the RVS at the DAO between 1988 and 1991 using a K star mask (DAO-RVS).  
These are all listed in Table~\ref{tab_22272RV}.
In addition, we have available 72 CORAVEL observations of IRAS 22272+5435 published by \citet{zacs09} obtained with the Moletai Observatory telescope and with the 1.5 m Steward Observatory telescope on Mt. Lemmon.
All of the observations have an internal precision of $\le$ 1.0 km~s$^{-1}$.

Special care was called for in combining these different velocity data sets obtained with different methods.  While all are recognized as giving precise velocities, we wanted to be careful to investigate and correct for any peculiar systematic effects.  It is known, for example, that a systematic offset occurs for CORAVEL data of red stars of late spectral types \citep{udry99}.
This is discussed in the following two sections.

\placetable{tab_22223RV}

\placetable{tab_22272RV}

\section{VARIABILITY STUDY OF IRAS 22272+5435}
\label{22272}

\subsection{Light Curve Study}

Our earlier{\it V} and {\it R$_C$} light curves of IRAS 22272+5435 from 1994 through 2007 have been described in Paper I.  The light varies in a cyclical pattern with the appearance of a changing amplitude.  A period of $\sim$130 days was found from the light and color curves, with a seasonal {\it V} amplitude that varies from 0.49 mag (1998, 2007) to 0.22 mag (1995).  The changing amplitude of the cyclical variation resembles a beat period and suggests a second period of similar value.  The object is redder when fainter.  This data set consists of 248 {\it V}, 189 {\it R$_C$}, and 132 ({\it V$-$R$_C$}) data points.  There is a gap in the {\it V} data from 2000 through 2002.  The light and color curves are displayed in that paper and the availability of the data is described there.     

\begin{figure}\figurenum{1}\epsscale{1.10} 
\plotone{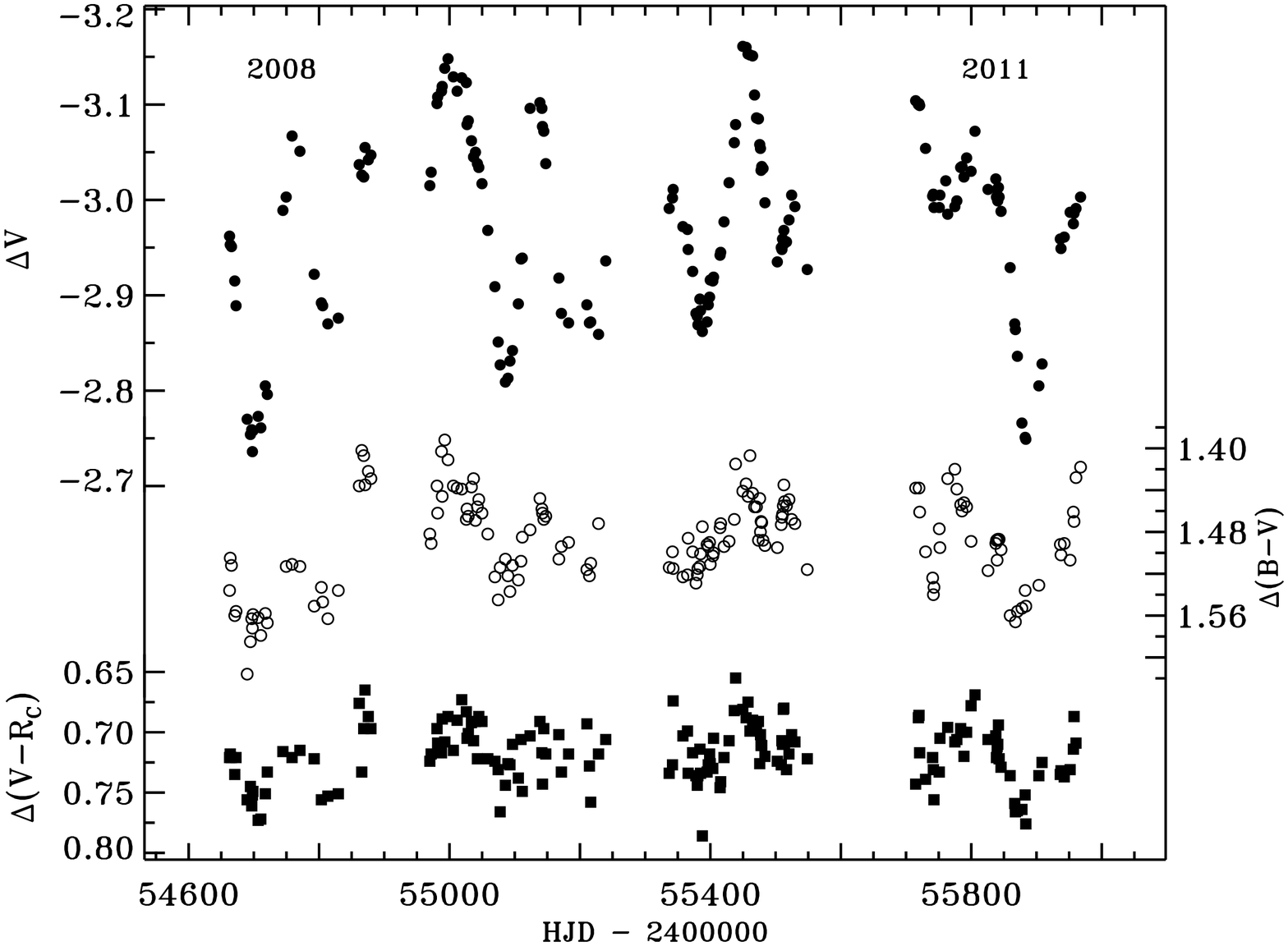}
\caption{Differential {\it V} light curve and ({\it B$-$V}) and ({\it V$-$R}$_C$) color curves of IRAS 22272+5435 from the 2008 through 2011 seasons.
\label{22272_lc_new}}
\epsscale{1.0}
\end{figure}

\begin{figure}\figurenum{2}\epsscale{0.75} 
\plotone{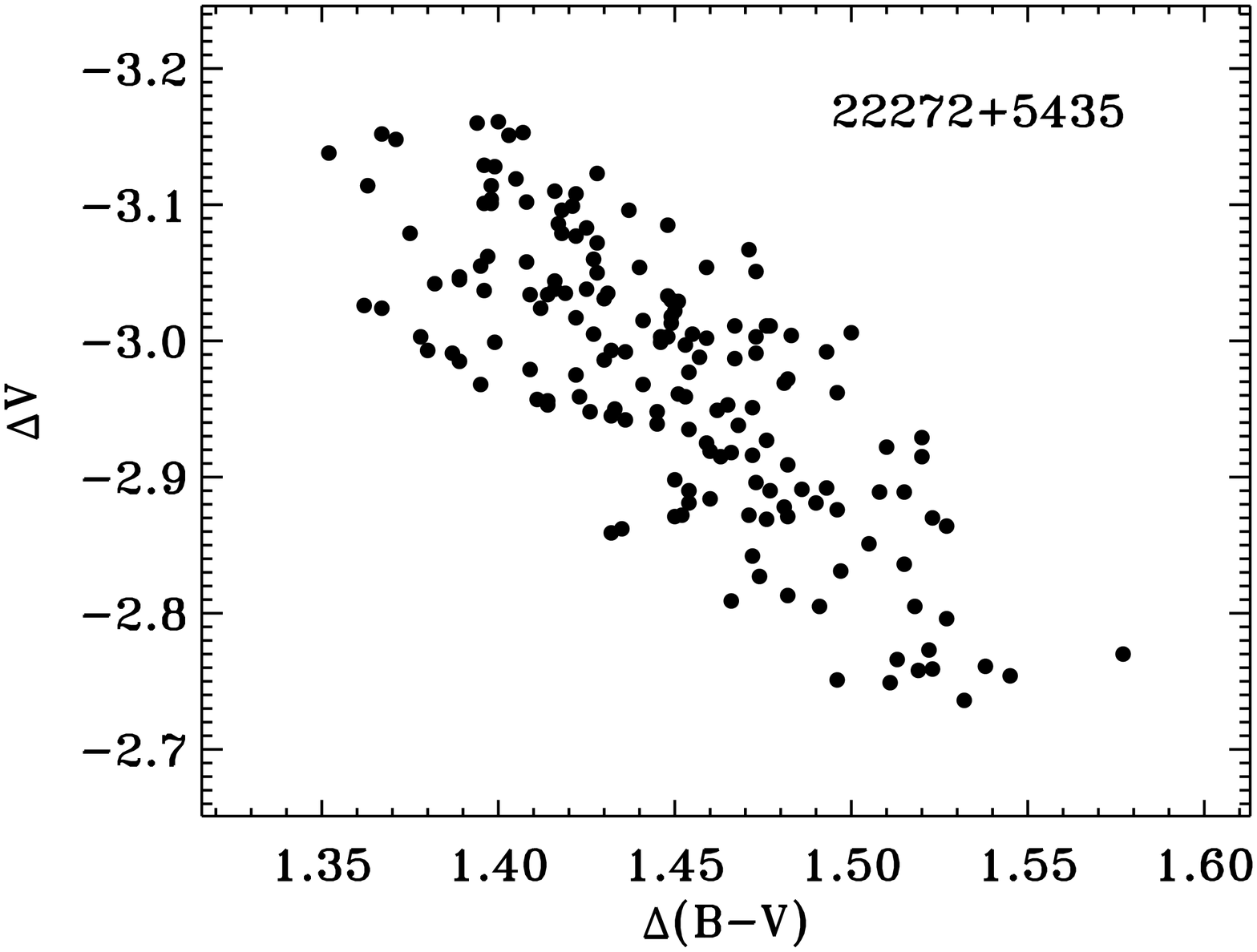}
\plotone{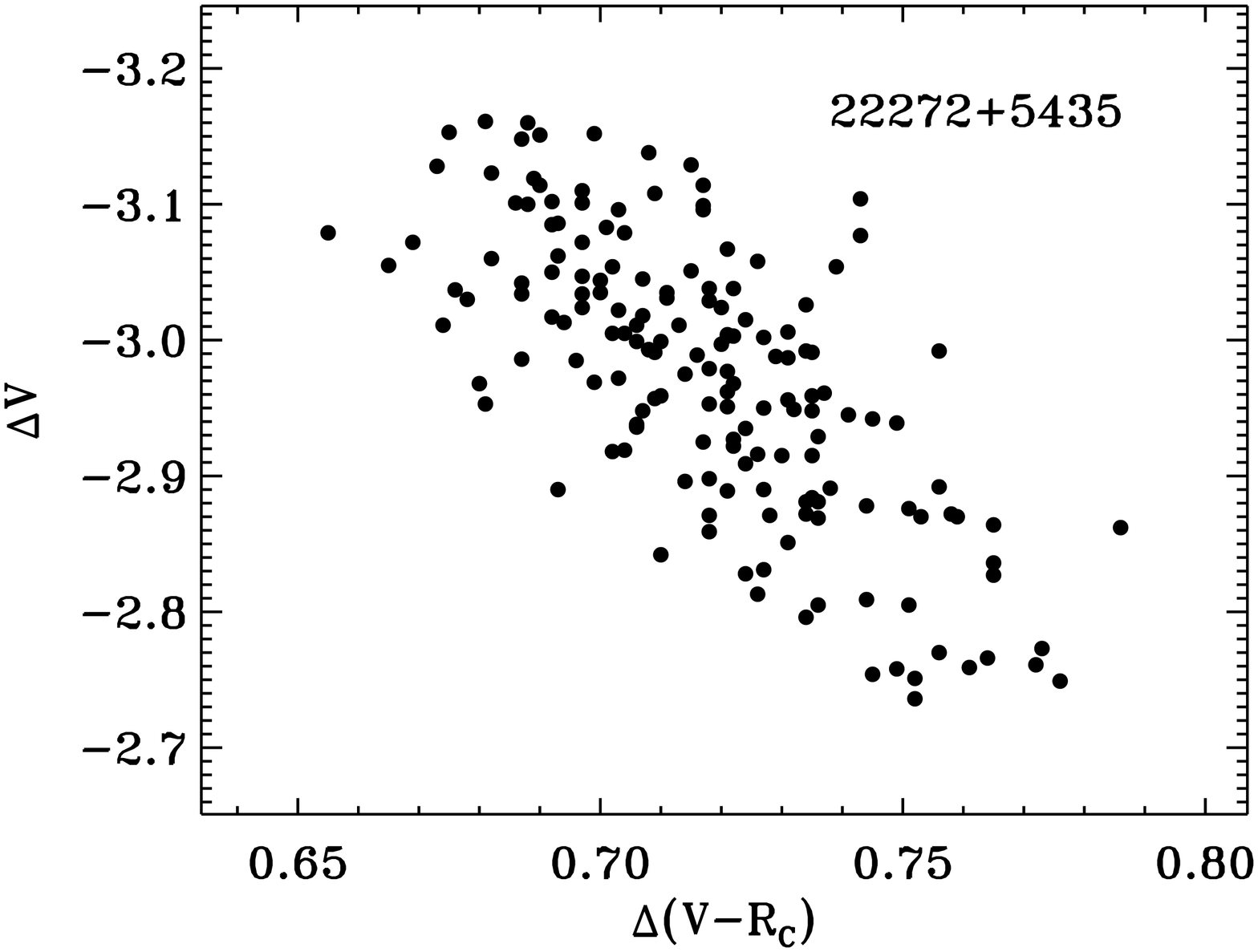}
\caption{Brightness versus color curves for the new observations of IRAS 22272+5435.  They clearly show the correlation of color with brightness; the object is redder when fainter.
\label{22272_cc_new}}
\epsscale{1.0}
\end{figure}

Our new {\it BVR$_C$} light curves from 2008 to 2011 show similar cyclical variations with an average {\it V} amplitude of $\sim$0.34 mag, a larger amplitude in {\it B} ($\sim$0.41 mag), and a slightly smaller amplitude in {\it R$_C$} ($\sim$0.31).  These new light and color curves are shown in Figure~\ref{22272_lc_new}. 
The depths of the minima vary quite a bit, from being similar in 2009 to differing by a factor of about three in 2011.  The pattern differs from the regular pattern of deep minimum followed by shallow minimum seen in RV Tauri variables.
The color curves mimic the cyclical variations in the light curve, with the object again seen to be redder when fainter.
This can also be seen clearly in Figure~\ref{22272_cc_new}, a plot of brightness versus color.
These data show good continuity with our older data; the comparison star data appear to be consistent and the light and color curves show similar average values and ranges of variability.

\placefigure{22272_lc_new}

\placefigure{22272_cc_new}

\citet{ark93,ark00} have published photometric studies of IRAS 22272+5435 from 1991$-$1999 and publically made available their {\it UBV} data, which consist of 89 observations in each filter.  
There is thus overlap between the two data sets from 1994$-$1999.

To investigate further the variability in the system and to search for multiple periods or changing periods, we have combined our two sets of {\it V} observations together with the {\it V} data of \citet{ark93,ark00}.  
In Paper I we published differential {\it V} magnitudes with respect to a comparison star, as we have done here, although using a different comparison star.  Using the standard magnitudes of the comparison stars ({\it V}(C$_{\rm 1,new}$)=11.62; {\it V}(C$_{\rm 1,old}$)=11.16), we have formed {\it V} magnitudes for our observations of IRAS 22272+5435.
Arkhipova et al. published their light curve data in the form of {\it V} magnitudes.  

\begin{figure}\figurenum{3}\epsscale{1.10} 
\plotone{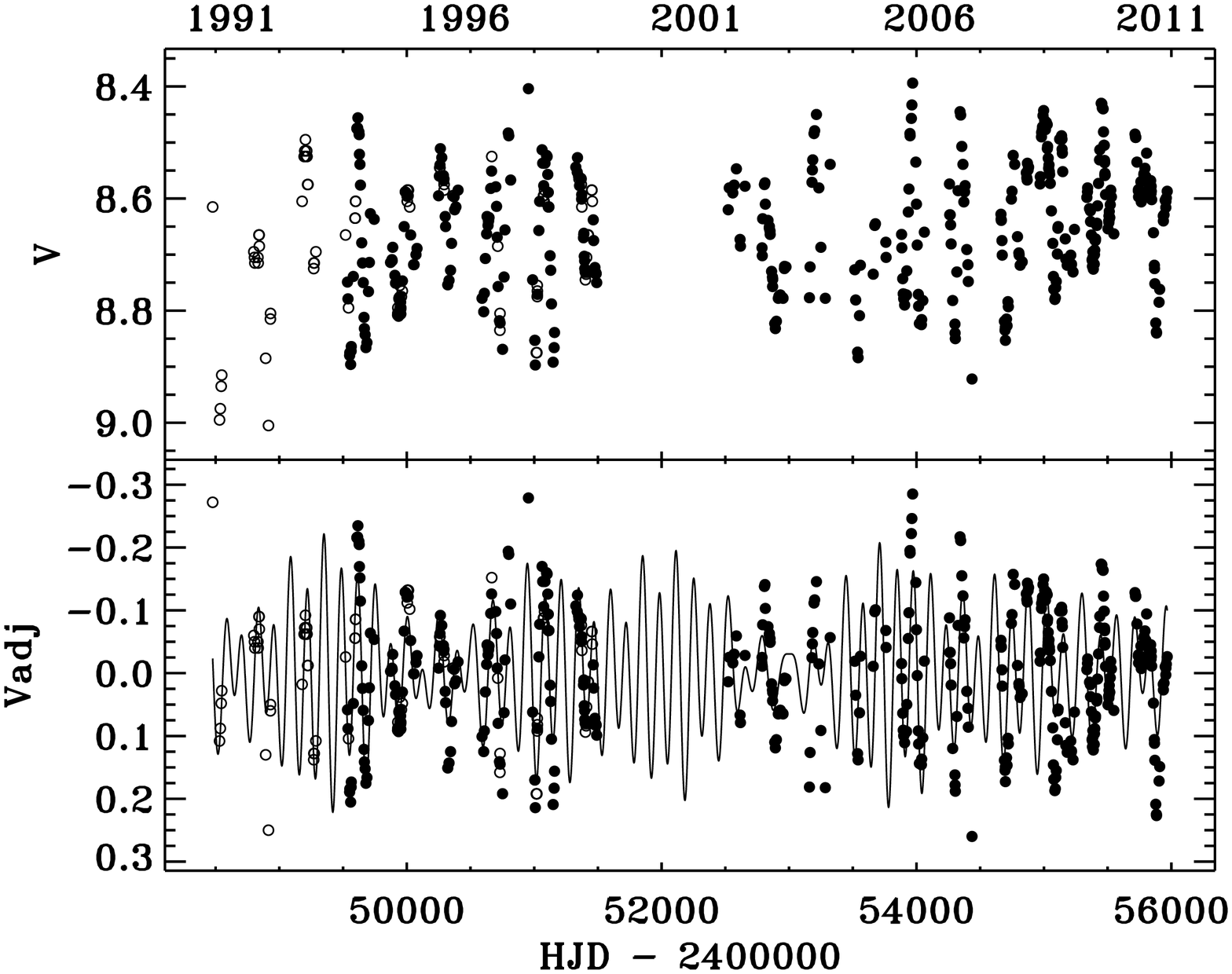}
\caption{Top: The combined {\it V} light curve of IRAS 22272+5435 from the 1991 to 2011 seasons.  
Bottom: The seasonally-adjusted, combined {\it V} light curve fitted with the four periods and amplitudes of the periodogram analysis.
The filled circles are the data from the VUO and the open circles from \citet{ark93,ark00}.
\label{22272_V-lc_all}}
\epsscale{1.0}
\end{figure}

An initial combination of the data revealed a systematic offset when plotted, with the data of \citet{ark93,ark00} brighter than ours.
We then compared the data on nights in which we both observed the object (nine nights in 1995$-$1998), and found an offset of 0.105 $\pm$ 0.006 mag.
Several factors could contribute to such an offset, such as a variation in the comparison stars or inclusion of another star in the aperture.\footnote{
We investigated several factors which could contribute to an offset in the two data sets, 
beginning with a possible variation in the comparison star.
We have four standardized observations of our initial comparison star made over a 17-year
interval, and they agree to within $\pm$0.015 mag  and our comparison stars appear to be constant at the level of $\pm$0.015 mag over the 18 seasons of our observations.  This leads us to think that there is not a systematic zero-point error in our data.
The comparison star used by
Arkhipova et al., BD+54$\arcdeg$2793 (HD 235865), has {\it B$-$V} = 1.91 and is classified as M2~III (SIMBAD), 
so it is easy to suspect that it varies at some level.  
\citet{ark93} list a value of {\it V} = 8.54 while in SIMBAD it is listed at {\it V} = 8.60.  So the cause could be a variation in their comparison star.
They used a photoelectric photometer with a large aperture of 27$\arcsec$ diameter, which thus includes a star near IRAS 22272+5435; we measured that the effect of this would be to increase the brightness attributed to IRAS 22272+5435 of $\sim$0.015 mag.  Some combination of these factors and the uncertainty of $\pm$0.015 mag in our standard star measurements likely explains the systematic difference in brightness of the two data sets.}
For the purposes of this period study, we have added an offset of +0.105 mag to each of the
{\it V} magnitudes of \citet{ark93,ark00} to bring the two data sets into agreement.
The resulting combined {\it V} light curve is shown in Figure~\ref{22272_V-lc_all}.
The combined {\it V} light curve displays the variations in seasonal amplitude and in seasonal mean brightness seen previously in our data alone.  
The variations in seasonal amplitude are due to the presence of multiple pulsation periods, as will be discussed shortly, which can enhance or diminish the amplitude.  
The variations in mean seasonal brightness might be due to temporal changes in the circumstellar opacity, since the star is known to be surrounded by a dusty nebula.

\placefigure{22272_V-lc_all}

The combined {\it V} data set and the VUO {\it R$_C$} and ({\it V$-$R$_C$}) data sets were searched for periodicity using the program Period04 \citep{lenz05}.  We explored multiple periods and found at least four significant periods in most of the data sets.
A period was judged to be significant if the S/N amplitude ratio of the peak in the frequency power spectrum was $\ge$ 4 \citep{bre93}.
In light of the aforementioned seasonal variations in the mean light level, we began by comparing the results of a period analysis based on the observed data with one in which the observations of each season were normalized to the mean brightness of the entire data set.  
This was carried out for the combined {\it V} light curve.  
The results were that the dominant first period and also the second period were the same in both cases.  
The next two differed, but there was no evidence of a periodicity in the seasonal means.  Thus we will report the period study based on the seasonally-normalized {\it V}, {\it R$_C$}, and {\it V$-$R$_C$} data, with the assumption that this better represents the pulsational properties of the object. 
The results are as follows, and are tabulated in Table~\ref{22272_lc_per}.

\placetable{22272_lc_per}

\begin{figure}\figurenum{4}\epsscale{0.75} 
\plotone{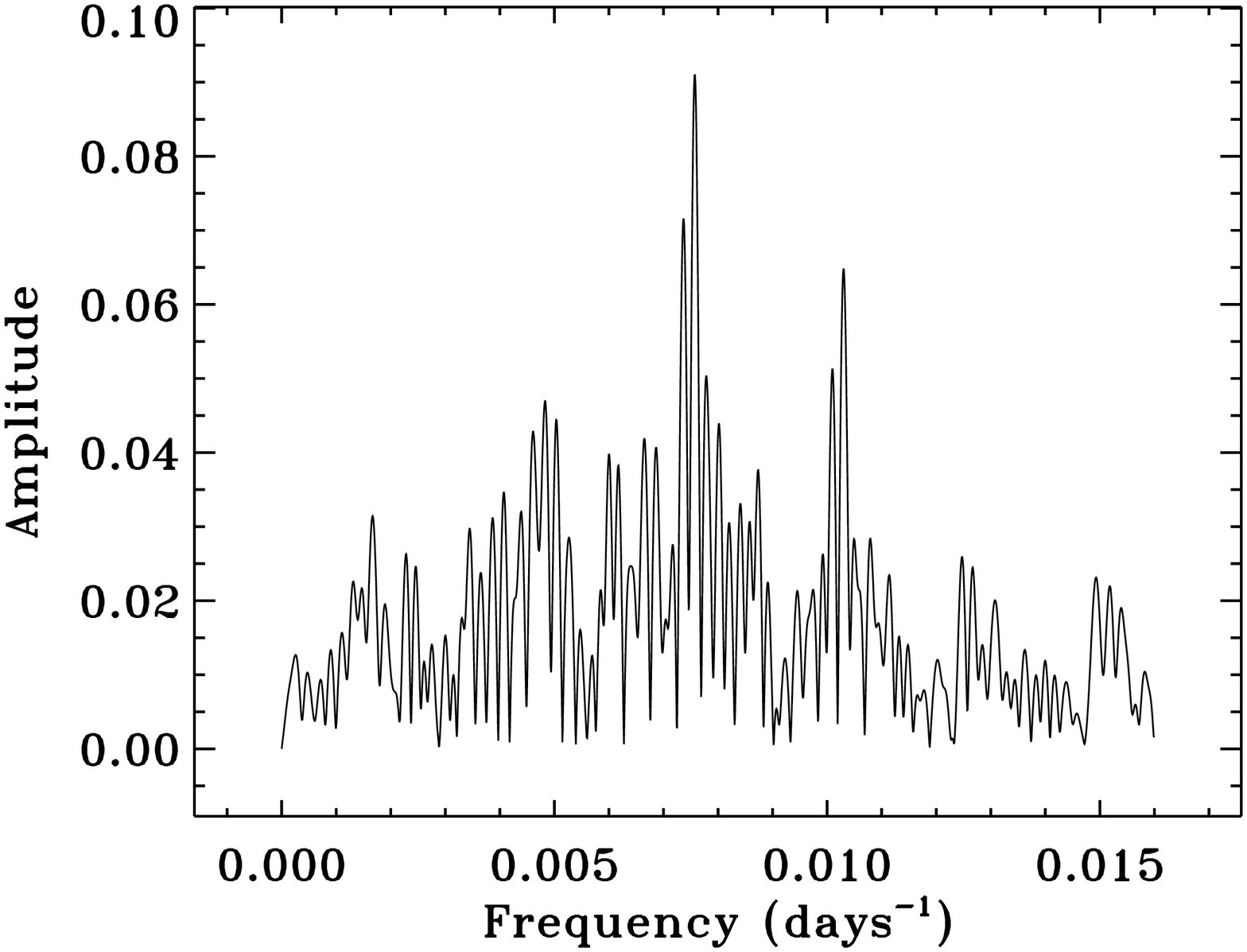}
\plotone{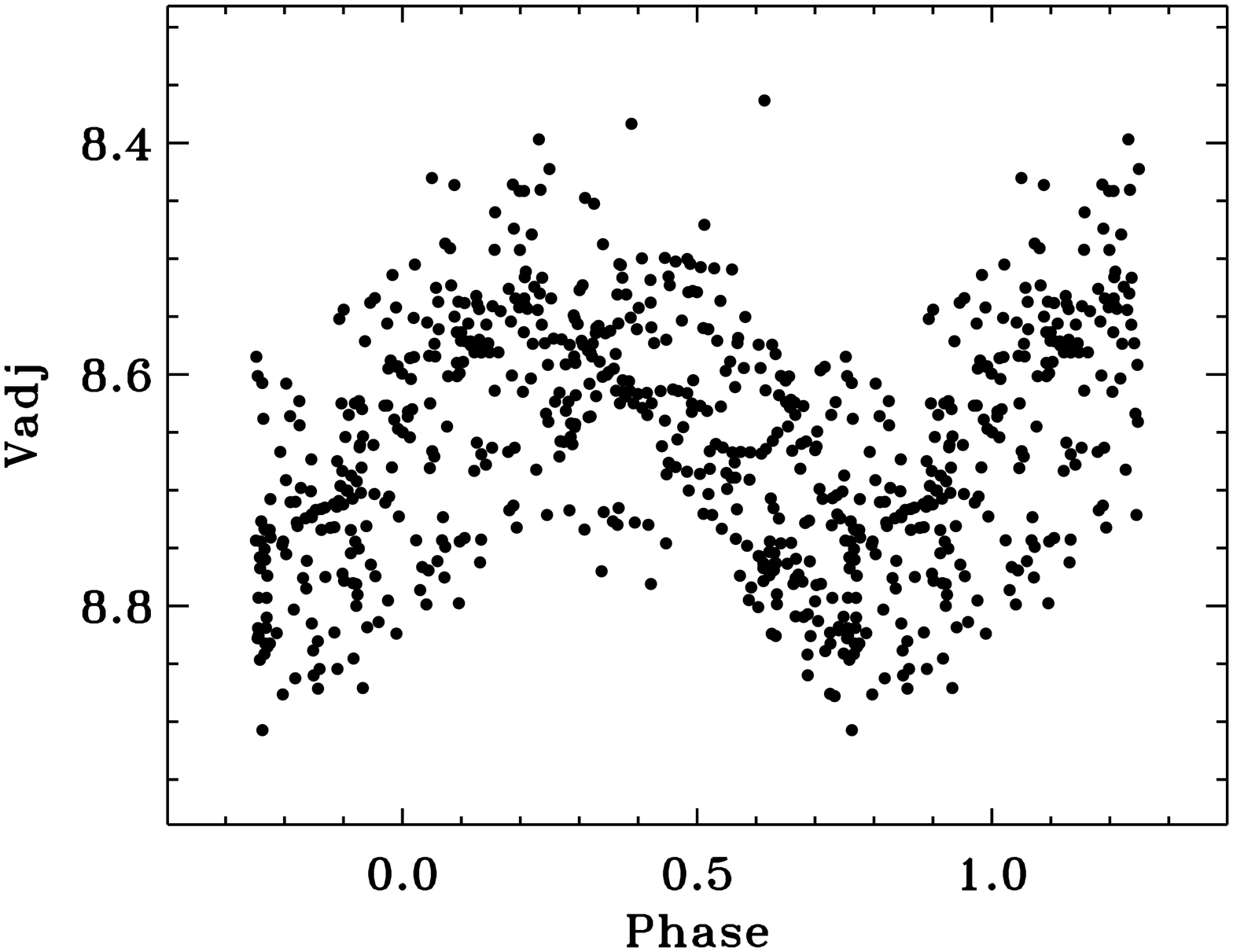}
\caption{Top: The frequency spectrum for the first period of the {\it V} light curve of IRAS 22272+5435.
Bottom: The phase plot of the seasonally-normalized data based on the frequency peak, P$_1$=131.9 d.
\label{22272_freqspec}}
\epsscale{1.0}
\end{figure}

{\it A. The combined {\it V} data set from 1991$-$2011}:
For the combined {\it V} data set, there is a dominant period of 131.9$\pm$0.1 days with a second period of 125.0$\pm$0.1 days. The combination of these two periods gives a reasonably good fit to the modulated light curve.
The ratio of these periods is P$_2$/P$_1$ = 0.95.  The power spectrum of the first period is shown in Figure~\ref{22272_freqspec}, along with a phase diagram from this single period.  Comparing the Fourier components of the fit with four periods to the light curve results in somewhat better agreement than does the fit with two; this fit is shown in the bottom panel of Figure~\ref{22272_V-lc_all}.  P$_4$ is somewhat uncertain and an almost equally good value is 155.9 days, which equals 2/3 of P$_4$.

\placefigure{22272_freqspec}

{\it B. The combined {\it V} data set from 1991$-$1999}:
The combined {\it V} data from the first nine years yield a similar value for P$_1$ but do not contain the period of 125 days seen as P$_2$ in the entire data set.

{\it C. The {\it V} data set of  the VUO from 2002$-$2011}:
The {\it V} data from the last ten years yield values for P$_1$ and P$_2$ that are similar to those found for the entire data set (set {\it A} above).  The values for P$_3$ and P$_4$ are $\sim$5$\%$ and $\sim$10$\%$ larger than those for the entire data set, respectively, with P$_4$ $\approx$ 2P$_2$.

{\it D. The {\it R$_C$} data set of  the VUO from 2002$-$2011}:
The {\it R$_C$} data from the last ten years were also investigated.  They yield values of P$_1$ and P$_2$ similar to those found for the corresponding {\it V} data set and for the entire {\it V} data set.  However, the values of P$_3$ and P$_4$ are different, and a value for P$_4$ of 222.6 days is almost as significant as that of 138.1 days as listed.  The amplitude of the {\it R$_C$} light curve is significantly smaller than the {\it V} light curve, and this makes especially the weaker periods less certain.

{\it E. The {\it R$_C$} data set of  the VUO from 1999$-$2011}:
The {\it R$_C$} data beginning in 1999, the first year in which we obtained a reasonable number of data points, were also examined.  This 13-year data set yielded values for the periods similar to those found for the slightly more restricted {\it R$_C$} data set from 2002$-$2011, with P$_1$ and P$_2$ close to those found for the combined {\it V} data set.

{\it F. The {\it V$-$R$_C$} data set of  the VUO from 2002$-$2011}:
The {\it V$-$R$_C$} data from the last ten years yield values for P$_1$ and P$_2$ that are close ($<$1$\%$ difference) to those found from the other data sets.
Thus they show clearly that a similar periodicity is found in the color curve as in the light curves.

These results show a consistent primary period of 131 days throughout this 20-year observing interval, with no indication of a secular change in the primary period.  The value of P$_2$ = 125 days, however, is not found in the 1991$-$1999 data set, although it is seen in all of the data sets from 1999$-$2011 and in the combined 1991$-$2011 {\it V} data set.  These values yield a ratio of P$_2$/P$_1$ = 0.95.

\begin{figure}\figurenum{5}\epsscale{1.10} 
\plotone{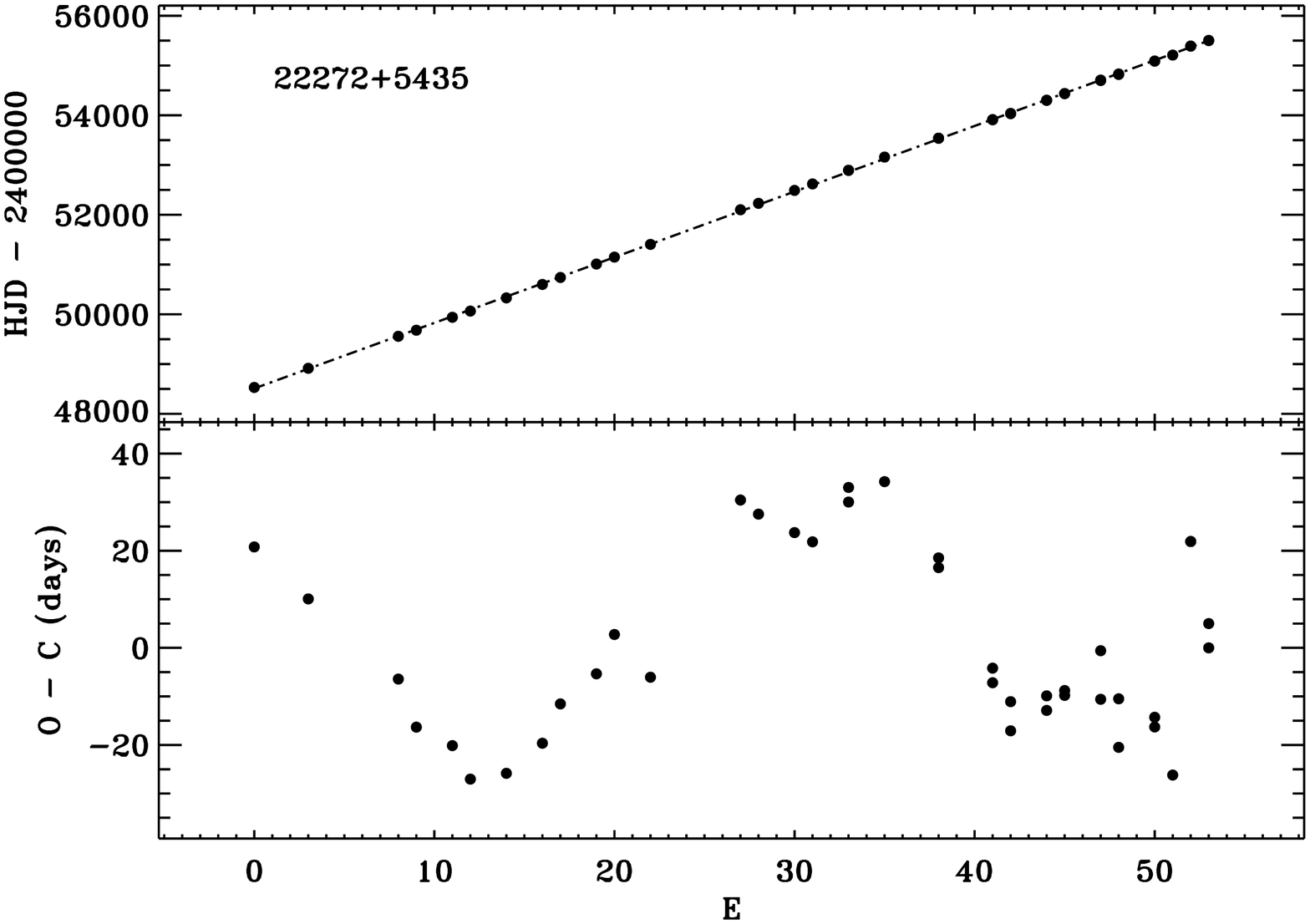}
\caption{Top: Observed time of minimum versus cycle count for the {\it V} and {\it R}$_C$ light curves of IRAS 22272+5435.    
The slope of the line is 131.9 day/cycle.
Bottom: The residuals of the time of minimum about the straight line fit.  
\label{22272_Tmin_E}}
\epsscale{1.0}
\end{figure}

To investigate further any evidence for a secular change in the period, we plot in Figure~\ref{22272_Tmin_E} cycle number versus Heliocentric Julian Date (HJD) for the minima of IRAS 22272+5435.  This is based on visual inspection of the VUO {\it V} and {\it R$_C$} light curves.  On this scale, the data points appear to fit a straight line, with a slope yielding a period of 131.9 days, the main period determined from the periodogram analysis.  Also plotted are the residuals from this straight-line fit.  They show the minima initially occurring later than predicted, but then gradually occurring earlier, as if the period were shortening, then later as if the period were lengthening, then occurring earlier.  This suggests a cyclical pattern with a period of $\sim$34 cycles or $\sim$4500 days (12.3 yrs), although future data are needed to confirm that the minima are occurring later than predicted by the linear ephemeris.  An obvious thought is that this is showing the longer-term beat period.  However, based on P$_1$ = 131.9 days and  P$_2$ = 125.0 days, the beat period is 2389 days, while the apparent period seen in the residuals is almost twice that value.
We compared these residuals with those derived from a linear fit to the times of minimum of the theoretical {\it V} light curve based on the four periods and their amplitudes, as shown in Figure \ref{22272_V-lc_all}.  This resulted in residuals of similar ({\it O$-$C}) range but which did not show the same pattern.

\placefigure{22272_Tmin_E}

\subsection{Radial Velocity Study}
\label{22272_rv}

As described earlier, we have carried out a long-term series of radial velocity observations of IRAS 22272+5435, initially from 1991$-$1995, and then resuming in 2005 and continuing to the present time.  
Recently, two of us published an intensive radial velocity study with observations from 2005$-$2007, especially from the 2006 season \citep{zacs09}.   

Initial examination of the more recent CORAVEL and DAO-CCD observations for IRAS 22272+5435 showed an apparent offset in the velocities observed in the same seasons.  
Thus we decided to explore this further by comparing the radial velocity measurements of IRAS 22272+5435 from the three observatories.
We compared the differences 
(a) in the average values of the velocities from each observatory for all nights,
(b) in the average values of the velocities only over the years in which they were observed in common at the different observatories, 
(c) in the individual velocities observed on common nights or successive nights at two different observatories, and 
(d) in the systemic velocities ({\it V$_{\rm 0}$}) when we were able to compute velocity solutions for the data from an individual observatory.
Each of these methods of comparison has its limitations:
(a) the yearly averages are not necessarily the same from year to year,
(b) and even within a year, the different distribution of observation dates may skew the averages from the different observatories.
Method (c) is the best in principle, but the number of nights in one case is small (6).
We don't find radial velocity solutions for all of the data sets alone (not for DAO-CCD), so method (d)  cannot be applied to each.
Comparing the different methods, we arrived at the following offsets, which were added to the different data sets to bring them to the same velocity system for IRAS 22272+5435, which we arbitrarily set to that of the DAO-CCD observations.  
The offsets are as follows: $-$1.4 km~s$^{-1}$ for the CORAVEL measurements and $-$0.2 km~s$^{-1}$ for the Hermes measurements.  The latter may or may not be real, but there is a real, significant difference between the CORAVEL measurements and the other two data sets.\footnote{
IRAS 22272+5435 has an usual spectrum, with strong lines due to s-process elements and very strong reddening of its continuum.  The three different instruments used for the 2005$-$2011 measurements each sample different regions of the spectrum and at different resolutions.  Thus, it is perhaps not surprising that there are zero-point differences in the measurements for IRAS 22272+5435, even though each of the different velocity systems is well calibrated.
This could be investigated further using the spectrum of IRAS 22272+5435 over the various wavelength regions and simulating the different detector resolutions and analysis methods.  However, we have chosen not to do so within the scope of this study.} 

We note that \citet{kloch09} have published four radial velocity observations and \citet{red02} one observation from 2000 of IRAS 22272+5435.
However, in light of the systematic differences that we have found among our three data sets, 
we have chosen not to include these few additional observations in our study.

\begin{figure}\figurenum{6}\epsscale{1.00} 
%\plotone{22272_RadialVelocity.eps}
\plotone{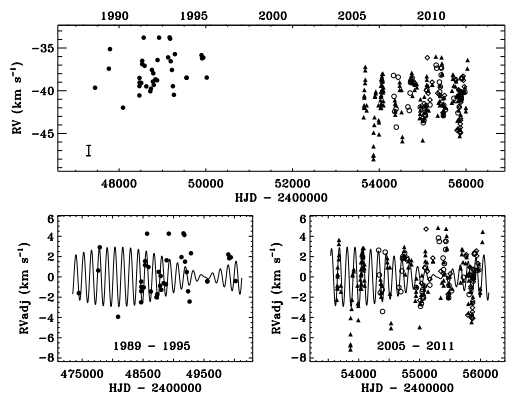}
\caption{Top: Radial velocity observation of IRAS 22272+5435 from 1989 to 2011, 
including the offsets to bring together the measurements from the different observatories.  
The symbols represent the observations from DAO-RVS (filled circle), DAO-CCD (open circle), CORAVEL (filled triangle), and Hermes (open diamond).
On the lower left is shown a sample error bar for the DAO and CORAVEL data; one for the Hermes data  would be smaller.
Bottom: Normalized radial velocities fitted with the first three periods and amplitudes from the radial velocity periodogram analysis.  The early DAO-RVS measurements have a zero-point adjustment of $-$2.80 km~s$^{-1}$ to normalize the data.  See the text for more details.
\label{22272_rv_all}}
\epsscale{1.0}
\end{figure}

We have combined these data sets with the above offsets to investigate the period and radial velocity properties of IRAS 22272+5435.  
This combined radial velocity curve is shown in Figure~\ref{22272_rv_all}.
One can see a variation in the radial velocities that reaches up to 10 km~s$^{-1}$ within a season (2006).  
This is the pulsational variation reported in the preliminary study of \citet{hri00} and the study of \citet{zacs09}.
It can be seen that there also exists a systemic difference between the early velocity measurements from 
1988$-$1995 and the later measurements.
This has been discussed by \citet{hri11} and attributed to a binary companion.  
Since our goal is to study the pulsational properties of the object, we have removed this difference by normalizing the earlier and later data by a zero-point adjustment to the earlier data ($-$2.8 km~s$^{-1}$).  
There also exist variations in the mean values from season to season, with, in the extreme cases, the measurements in 2006 extending to more negative values and those from 2010 extending to less negative values than the others.  While we saw an analogous effect in the light curves that we attributed to varying opacity in the circumstellar envelope, we have no similar explanation for this velocity change and have not carried out a seasonal normalization of the velocity curves as we did for the light curves.

\placefigure{22272_rv_all}

A periodogram analysis was carried out of this combined, normalized radial velocity data set.  
We analyzed the entire data set and also the earlier and later data separately.
The results are listed in Table~\ref{22272_vc_per}. 

\placetable{22272_vc_per}

{\it A. The combined radial velocity data set from 1988$-$2011:}  We found three significant periods: P$_1$ = 131.2$\pm$0.1 days, P$_2$ = 125.5$\pm$0.1 days, and a weaker P$_3$ = 66.8$\pm$0.1 days, a little more than half the value of P$_1$.
The values of P$_1$ and P$_2$ are very close to those determined from the combined {\it V} light curve.  
However, for  the light curve the period of $\sim$132 days was clearly more dominant while for the velocity curve the two periods are more similar in amplitude.  
The fit of the velocity curve with the Fourier components based on three periods gives a reasonably good fit to most of the data.  
This is shown in the bottom panels of Figure~\ref{22272_rv_all}.
However, the fits are not good in the regions where the  
sine curves interfere destructively, such as the fit to the 2009$-$2010 data. 
Also the fit is not good in 2011, where the two minima in the velocity curve are separated by $\sim$67 days, the value of P$_3$.    
Note, however, this period of $\sim$67 days is not found only in the 2011 and perhaps 2010 data; it was found as a secondary period in the 2005$-$2007 data of \citet{zacs09}.

{\it B. The combined radial velocity data set from 2005$-$2011:} 
For the later data set, one finds periods similar to those found from the entire velocity data set, with P$_1$ = 132.7$\pm$0.5 days the most dominant. However, when we try to vary all three periods simultaneously to achieve the best overall fit, they do not converge;
instead we find the two similar periods continue to move closer together and the amplitudes continue to increase.
The value of P = 67 days is again found and appears to at least partly derive from the 2011 velocities which have this smaller interval between two velocity minima.

{\it C. The combined radial velocity data set from 1988 to 1995:} This is a small data set, with only 39 observations, and the values are less certain, with P$_1$ = 126.2$\pm$1.0 days and P$_2$ = 148.5$\pm$1.7 days.  The first of these is a value similar to the value of P$_2$ in most of the velocity and light curve analyses.  The second is similar to the value of P$_2$ found in the analysis of the 1991$-$1999 {\it V} light curve, which covers a somewhat similar set of dates.

We also investigated the fit to the normalized radial velocity curve from 1988$-$2011 using the values of P$_1$ and P$_2$ fixed at those determined from the combined, normalized {\it V} light curve.  
The fit was good and a third significant period of 66.8 days was found, the same as found in set A above.
However, the fit is not quite as good as when the periods are determined from the velocity curve itself.

\subsection{Contemporaneous Light, Color, \& Velocity Curve Study}

\begin{figure}\figurenum{7}\epsscale{1.15} 
\plottwo{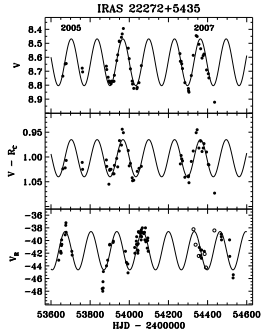}{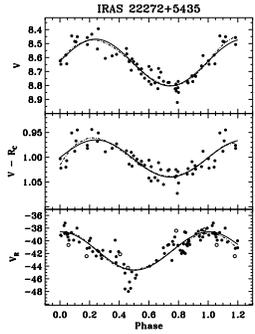}
\caption{(a, left panel) Contemporaneous {\it V}, ({\it V$-$R}$_C$), and {\it V$_R$} curves of IRAS 22272+5435 from 2005 to 2007, along with sine-curve fits based on P = 131.9 d.
(b, right panel) Phased plots (P = 131.9 d), along with sine curves (solid lines) and polynomial curves (dashed lines) fitted to the observations.  In the radial velocity panels, the filled circles are the data from \citet{zacs09} and the open circles from the DAO-CCD observations.
\label{22272_all_2005-07}}
\epsscale{1.0}
\end{figure}

The availability of these light and velocity observations from the same seasons affords the opportunity to compare contemporaneously the light, color, and velocity curves.
We begin with the large number of radial velocity observations obtained by \citet{zacs09} during the 2006 season, at a time when we also had good light and color curve observations.
To enlarge the sample, particularly to include the 2005 radial velocity and the 2007 light and color curve observations, we have combined the data from 2005$-$2007 beginning with the time of the radial velocity observations. 
These are shown in Figure~\ref{22272_all_2005-07}a.
For the subsequent radial velocity analysis, we have used only the \citet{zacs09} radial velocities  and have not included the six from DAO in 2007, although they are plotted in the figure and show good agreement.  
Sine curves were fitted to the data using the dominant period determined from the combined {\it V} light curve of 131.9 days and an arbitrary epoch of T$_0$ = 2,448,000.  
This yielded sine-curve values of semi-amplitude and phase, respectively, as follows $-$ {\it V}: 0.168 mag and 0.51, {\it V$-$R$_C$}: 0.038 mag and 0.51, and {\it V$_{\rm R}$}: 3.02 km s$^{-1}$ and 0.25, with uncertainties in the phase of $\pm$0.01.
Curves fitted with these parameters are overplotted on the observations in Figure~\ref{22272_all_2005-07}a, and they agree well with the observations.
One can immediately see that the light and color curves are exactly in phase according to a sine-curve fit, being exactly brightest when bluest.  
The radial velocity differs by 0.26 in phase, approximately a quarter of a cycle, being at the average velocity and minimum size when the star is brightest.

\placefigure{22272_all_2005-07}

The data are folded on this period and displayed during one cycle in Figure~\ref{22272_all_2005-07}b.
In this figure, in addition to using sine curves, we also fit the phased data with fourth-order polynomial curves.
These show that the velocity curve is essentially a sine curve, while the light and color curves appear to peak a bit earlier in phase ($\sim$0.02) and have a little more gradual decline to minimum  and a little steeper rise to maximum than does a sine curve.  
Given the dispersion of the observations about the lines, it can be seen that the sine curves represent the observations reasonably well.

\begin{figure}\figurenum{8}\epsscale{0.80} 
\plotone{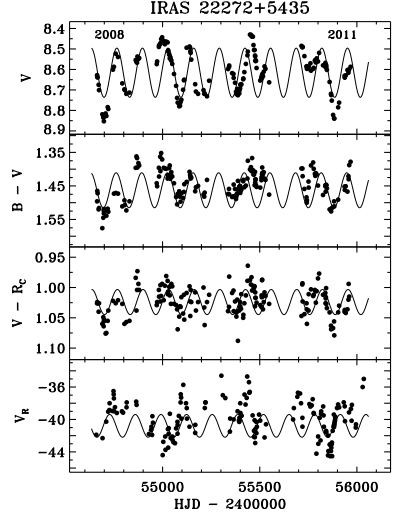}
\caption{Contemporaneous {\it V}, ({\it B$-$V}), ({\it V$-$R}$_C$), and {\it V$_{\rm R}$} curves of IRAS 22272+5435 from 2008 to 2011, along with sine-curve fits based on P = 131.9 d
and the amplitudes and phases derived from the 2008$-$2009 data only.  
\label{22272_all_2008-11}}
\epsscale{1.0}
\end{figure}

We similarly examined the 2008$-$2011 observations.  They are shown in Figure~\ref{22272_all_2008-11}.
The light and color curves have been displayed previously in Figure~\ref{22272_lc_new} and show considerable variation in the relative depths of minimum light and the mean levels in the different seasons.  
We have now normalized the seasonal light curves (but not the color curves) to their average values.
As mentioned earlier, the 2011 radial velocity observations display a period of $\sim$67 days, about half of the period found from the rest of the data.  This can be seen clearly in this figure.  
Sine curves were fitted to these observations with P = 131.9 days and the same epoch as above. 
We find that the fit to the {\it V} light curve is not very good when we include all four years; this is partly due to the large variations in the depths of minimum.  Similarly, the color curves are not fit well for all four years nor is the velocity curve.  Examining them over smaller intervals, the best fit is found when restricted to the 2008$-$2009 interval.
For the 2008$-$2009 observations, we find sine-curve values of semi-amplitude and phase, respectively,  as follows $-$ {\it V}: 0.120 mag and 0.55, {\it B$-$V}: 0.052 mag and 0.58, {\it V$-$R}: 0.021 mag and 0.53, and {\it V$_{\rm R}$}: 1.40 km s$^{-1}$ and 0.33.  The uncertainties in the phases range from 0.02$-$0.03.  The light and color curves appear to again be in phase and the radial velocity curves differs by 0.22 in phase, again about a quarter of a cycle; thus the phasing of the different curves agrees with the results found earlier for the 2005$-$2007 curves. 
We have plotted the sine curves with these parameters, derived from the 2008$-$2009 observations, through all of the 2008$-$2011 observations shown in Figure~\ref{22272_all_2008-11}.
It can be seen that the fit is not so good as found for the 2005$-$2007 observations, when the variations appeared to be less complex.
In particular, one can see the poor agreement to the period in the 2011 and perhaps the 2010 radial velocity data.  We will examine this in a future study in which we also include additional spectroscopic observations.  

\placefigure{22272_all_2008-11}

\section{VARIABILITY STUDY OF IRAS 22223+4327}
\label{22223}

\subsection{Light Curve Study}

Our earlier {\it V} and {\it R} light curves of IRAS 22223+4327 from 1994 through 2007 have also been described in Paper I. 
The light curves show a general decrease in brightness from 1994 to 2007, amounting to $\sim$0.16 mag in {\it V}.  Superimposed on this is a cyclical variation seen in the seasonal light curves.  The {\it V} amplitude of this cyclical variation changes from 0.21 mag (1997) to 0.09 mag (2002, 2006); this is about half of the corresponding range seen in IRAS 22272+5435.  
The object is also redder when fainter.  A period of $\sim$88 days is found in the light and color curves.   That data set consists of 267 {\it V}, 212 {\it R$_C$}, and 160 ({\it V$-$R$_C$}) data points, with a gap in the {\it V} data from 2000 through 2002.  The light and color curves are displayed in that paper and the data are made available.     

\begin{figure}\figurenum{9}\epsscale{1.00} 
\plotone{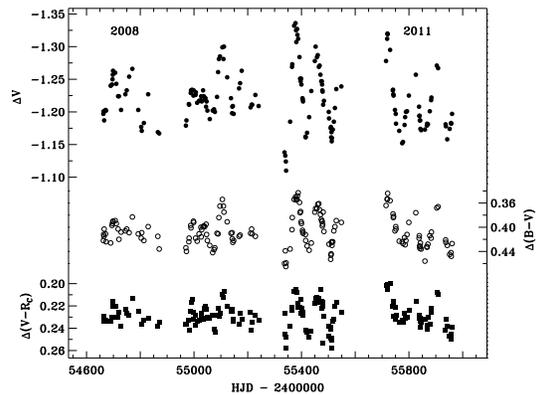}
\caption{Differential {\it V} light curve and ({\it B$-$V}) and ({\it V$-$R}$_C$) color curves of IRAS 22223+4327 from 2008 to 2011.\label{22223_lc_new}}
\epsscale{1.0}
\end{figure}

\begin{figure}\figurenum{10}\epsscale{0.75} 
\plotone{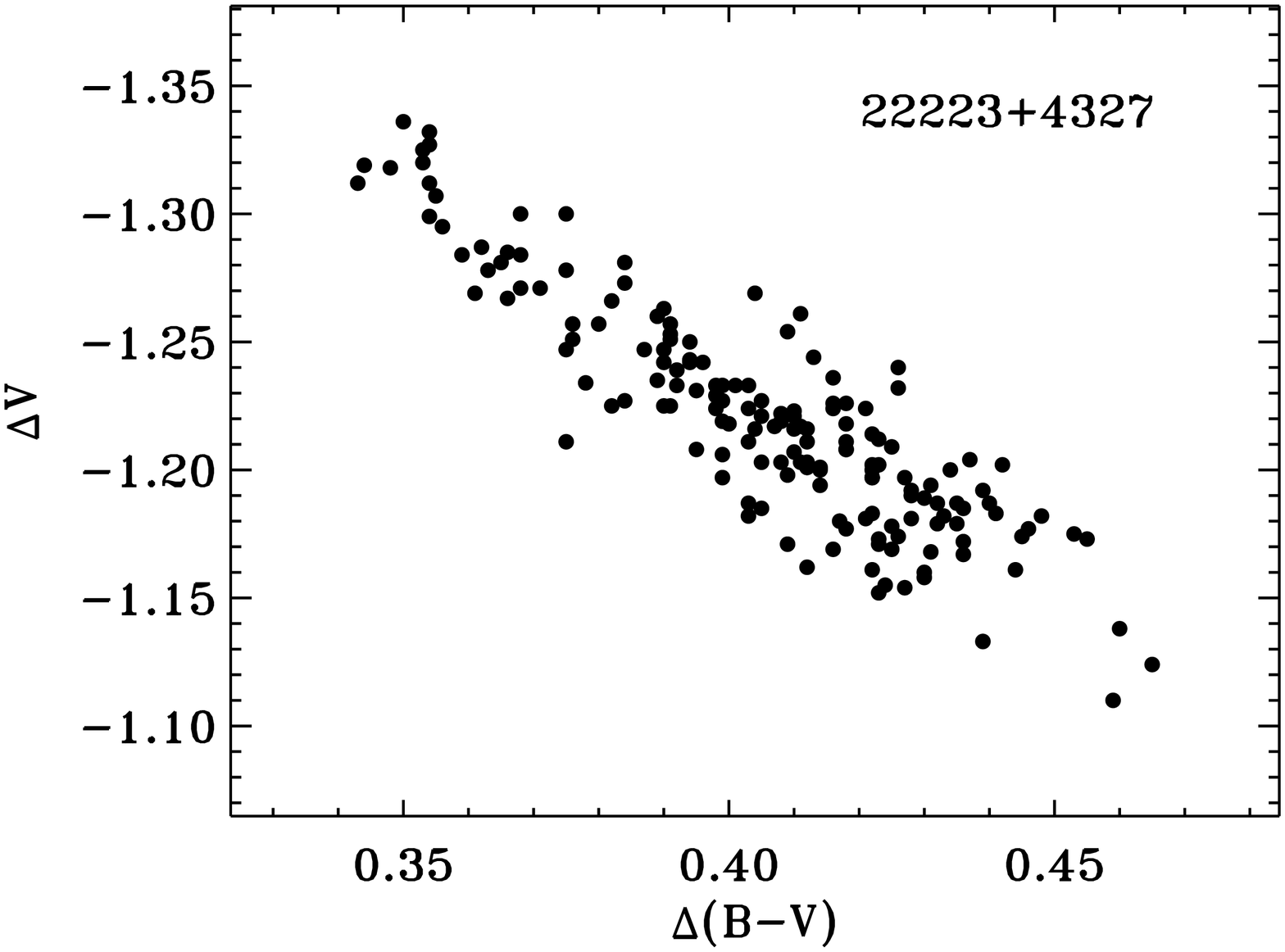}
\plotone{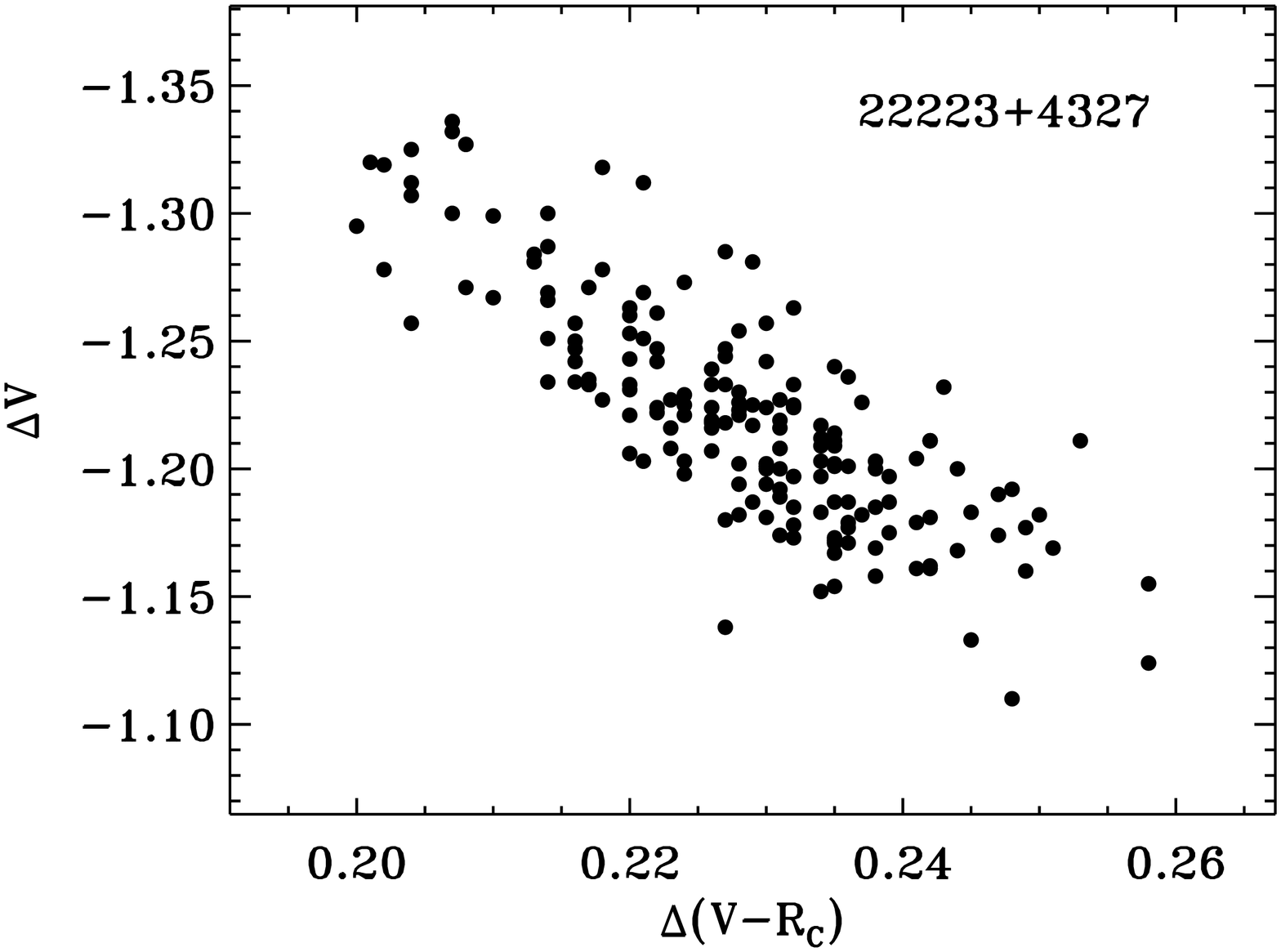}
\caption{Brightness versus color curves for our new 2008$-$2011 observations of IRAS 22223+4327.  They clearly show the correlation of color with brightness; the object is redder when fainter.
\label{22223_cc_new}}
\epsscale{1.0}
\end{figure}

Our new {\it BVR$_C$} light curves also show cyclical variations with a large amplitude change that ranges in {\it V} from 0.23 mag (2010) to 0.10 mag (2008).  The amplitude is larger in {\it B}, ranging from 0.31 to 0.14 mag, and slightly smaller in {\it R$_C$}, 0.19 to 0.10 mag.
However, the light curves no longer show a trend of decreasing brightness; rather, the median levels are approximately the same in each year.
Our newer and older observations for IRAS 22223+4327 appear to be consistent with each other based on the comparison star data and the color curve data.
These new light and color curves are shown in Figure~\ref{22223_lc_new}. 
The object is again seen to be redder when fainter, as seen in the color curves in  Figure~\ref{22223_lc_new} and as shown explicitly in the brightness versus color curves in  Figure~\ref{22223_cc_new}.

\placefigure{22223_lc_new}

\placefigure{22223_cc_new}

\citet{ark03,ark11} have published photometric studies of IRAS 22223+4327 from 1999$-$2002, consisting of 56 observations in each of the {\it UBV} filters, and from 2003$-$2010, consisting of 134 observations in each filter, respectively.
They also find a cyclical variation, with a maximum range of 0.25 mag ({\it V}), and period values of 85.5$\pm$1 days and 90.9 $\pm$1 days.  Their observations also show the object to be redder when fainter.  

\begin{figure}\figurenum{11}\epsscale{1.10} 
\plotone{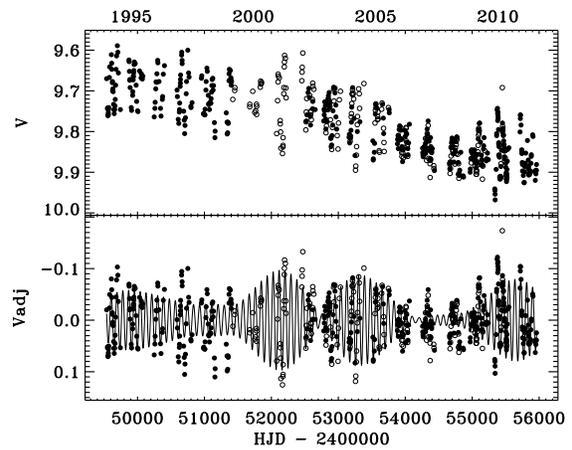}
\caption{Top: The combined {\it V} light curve of IRAS 22223+4327 from 1994 to 2011.  
Bottom: The combined {\it V} light curve fitted with the first four periods and amplitudes of the periodogram analysis.  The general trend of decreasing brightness was first removed using a fourth-order 
polynomial fit.
The filled circles are the data from the VUO and the open circles from \citet{ark03,ark11}.
\label{22223_V-lc_all}}
\epsscale{1.0}
\end{figure}

The {\it V} observations of ours (1994$-$2011) and those of \citet{ark03,ark11} were combined to study the period and light curve variations.  The two sets nicely complement each other and fill in the gap in the Paper I 
{\it V} light curve from 2000$-$2002.  
Our {\it V} observations are listed as differential magnitudes and were combined with the standard magnitude of the comparison star (V(C$_1$)=11.08) to give the {\it V} light curve.
When combining the two data sets, we investigated and found a small offset in brightness, with the \citet{ark03,ark11} data set brighter in the regions of overlap by 0.019 mag.\footnote{Note that the offset is based on 48 measurements of the difference between our standard magnitudes and those of \citet{ark03,ark11}, observed on the same night or with a difference one day between them.  This difference can be compared with the internal difference within our own data set, where we have observations on 49 nights in which we have data observed one night apart, and for which we found a difference of 0.002 mag; similar internal comparison of the data of Arkhipova et al. show a difference of $-$0.004 mag on 12 nights.}  
For the purposes of this period and light curve study, we have added an offset of +0.020 mag to each of the
{\it V} magnitudes of \citet{ark03,ark11}.
The resulting combined {\it V} light curve is shown in Figure~\ref{22223_V-lc_all}.

\placefigure{22223_V-lc_all}

This combined {\it V} light curve of IRAS 22223+4327 covers the interval 1994$-$2011.  As noted earlier, the light curves show a general decrease in brightness.  From 1994 to 2002, the {\it V} light curve decreased by 0.06 mag.  The decrease was more rapid from 2002 to 2008, 0.09 mag or $\sim$0.015 mag/yr.  From 2008 to 2011 the average light levels are approximately constant, indicating a leveling off of the decline.  This is also seen in our combined {\it R$_C$} light curve from 1995 to 2011.  We suggested earlier \citep{hri10} that this might be due to an increase in line-of-site dust extinction during this time.  However, the ({\it V$-$R$_C$}) color is constant over the entire interval of 1995$-$2011, with no evidence for increased reddening with decreased brightness; \citet{ark11} make similar comments about the constancy of the ({\it B$-$V}) and ({\it U$-$B}) color indices.  Thus, if the dimming is due to dust extinction, then the scattering must be gray, which would suggest relatively large ($\ge$1 $\mu$m) dust particles. 

We have proceeded to carry out a period analysis on the light curves of IRAS 22223+4327 similar to what we did with IRAS 22272+5435.  In this case, to account for the decrease in brightness, the {\it V} and {\it R$_C$} light curves were fit with a fourth-order polynomial, which was then used to remove the general monotonic decrease. 
The relatively large variation in the seasonal amplitudes suggests at least the existence of two periods in the data.
The results are as follows, and are tabulated in Table~\ref{22223_lc_per}. 

\placetable{22223_lc_per}

\begin{figure}\figurenum{12}\epsscale{0.75} 
\plotone{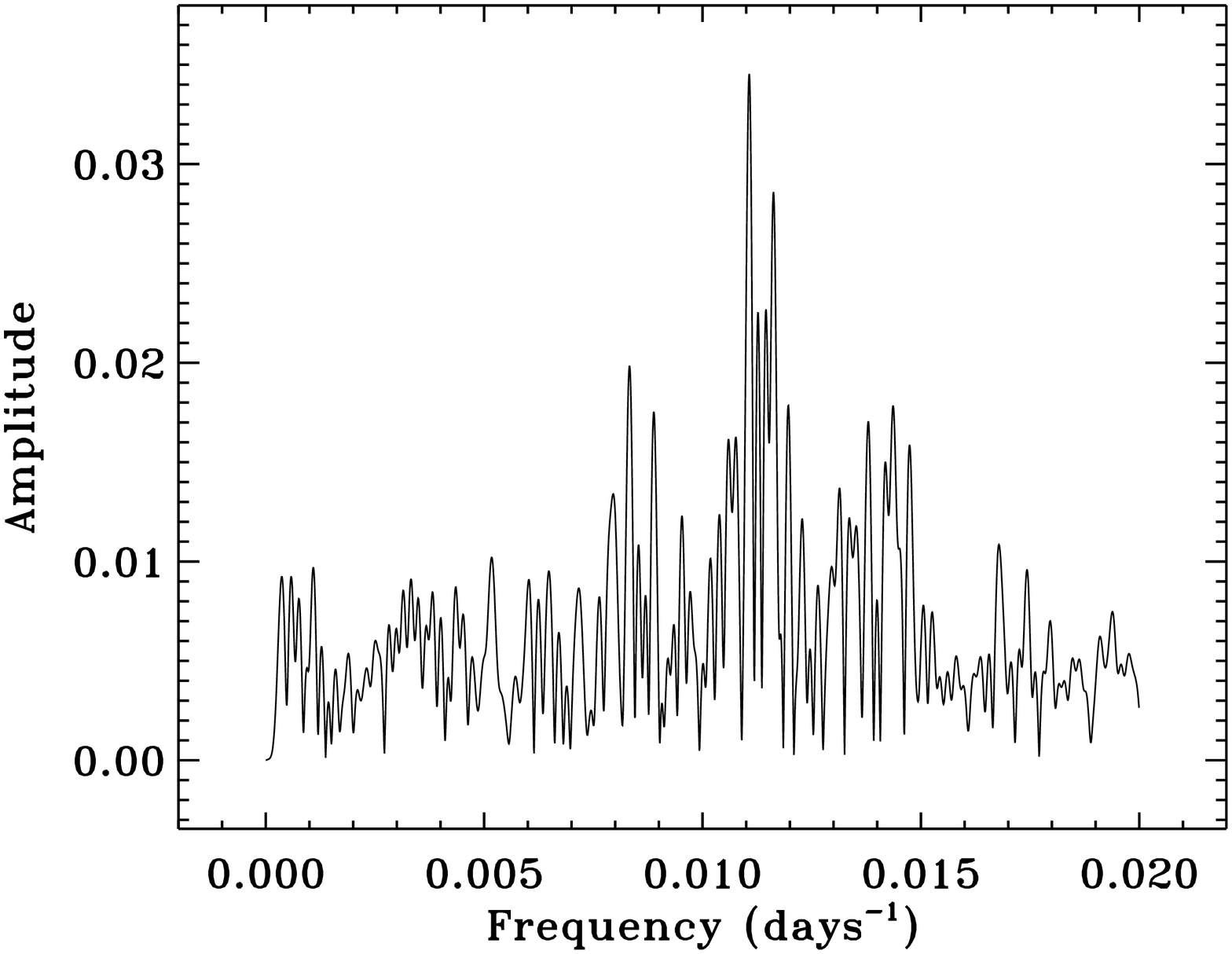}
\plotone{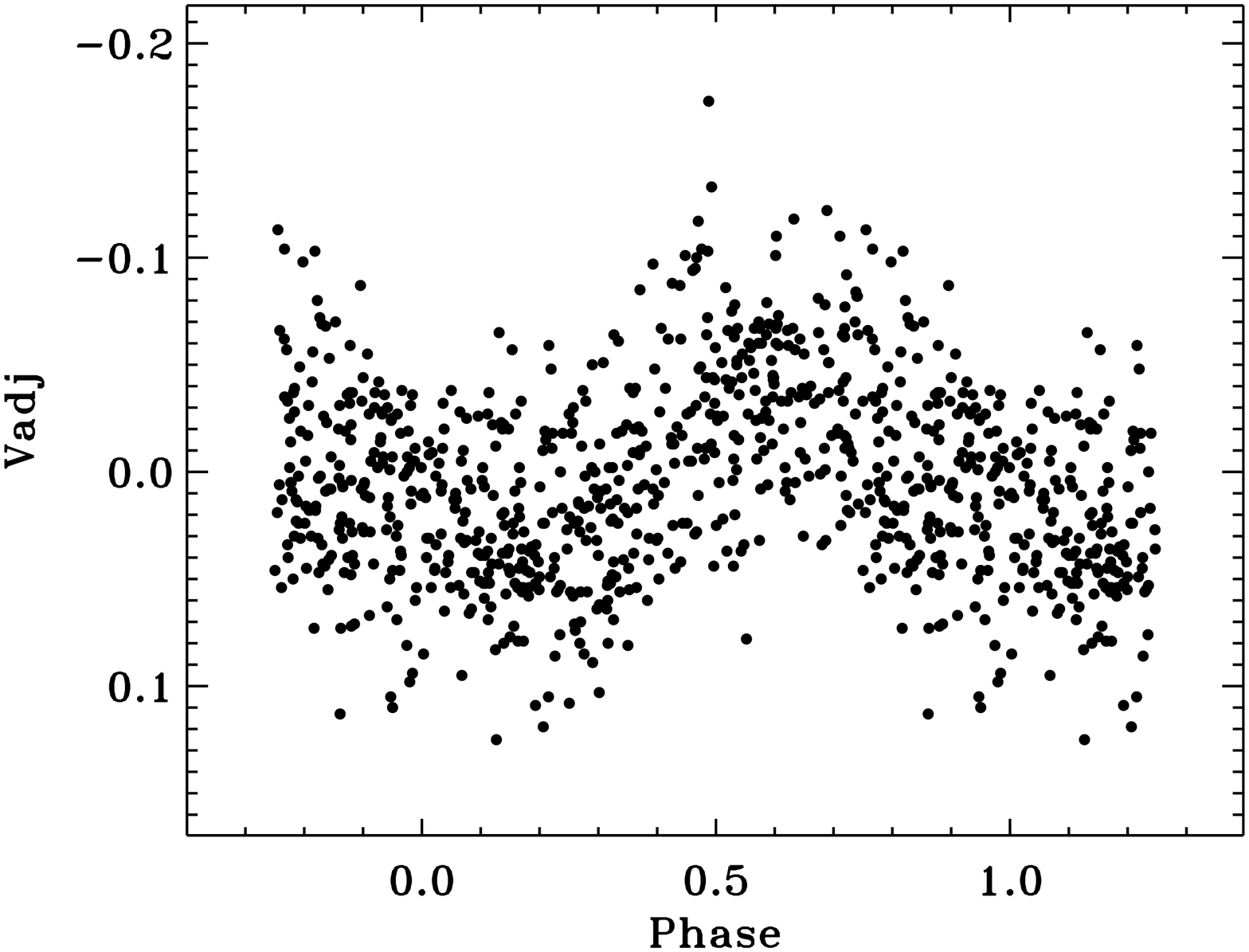}
\caption{Top: The frequency spectrum for the first period of the {\it V} light curve of IRAS 22223+4327.
Bottom: The phase plot of the {\it V} light curve based on the frequency peak of P$_1$=90.5 d.
These both use the adjusted data following the removal the long-term trend by a fourth-order 
polynomial fit.
\label{22223_freqspec}}
\epsscale{1.0}
\end{figure}

{\it A. The combined {\it V} data set from 1994$-$2011}:
For the combined {\it V} data set, there is a dominant period of 90.5$\pm$0.1 days with a second period of 85.8$\pm$0.1 days. 
These values are very similar to those found by \citet{ark11}, although we find the larger of the two to be the clearly dominant one.
The ratio of the periods is P$_2$/P$_1$ = 0.95, the same as found for IRAS 22272+5435.  The power spectrum of the first period is shown in Figure~\ref{22223_freqspec}, along with a phase diagram from this single period.  Comparing the Fourier components of the fit with two periods to the light curve results in reasonably good agreement, and this is improved with a fit to four periods, which is shown in the bottom panel of Figure~\ref{22223_V-lc_all}. 
While the constructive interference of several curves with different periods helps to fit the large amplitude seasons, the fit is poor, however, where the curves interfere destructively, such as in 2008 and 2009.

\placefigure{22223_freqspec}

{\it B. The combined {\it V} data set from 1994$-$2002}:
The combined {\it V} data set from the first nine years yields values that are similar to the periods determined from the entire combined {\it V} light curve, with P$_1$, P$_2$, and P$_4$ slightly smaller and P$_3$ slightly larger.  The first period is noticeably dominant in this data set; P$_4$ is barely significant.

{\it C. The combined {\it V} data set from 2003$-$2011}:
The {\it V} data set from the last nine years yields P$_1$ = 86.7$\pm$0.1 days, a value that is similar to P$_2$ of the combined data set, and P$_2$ = 89.5$\pm$0.1 days, a value that is similar P$_1$ of the combined data set (differing at the 1$\%$ level).  
These are similar to the values found by \citet{ark11} for their 2003$-$2010 data.
The amplitudes of P$_1$ and P$_2$ in the 2003$-$2011 data set are relatively similar, and neither one is dominant, in contrast to the the 1994$-$2002 {\it V} results.  There is no significant P$_4$ in this data set.  

{\it D. The {\it R$_C$} data set of  the VUO from 2003$-$2011}:
The {\it R$_C$} data set from the last nine years was also investigated.  It yields values for P$_1$ and P$_2$ that are very similar to those found from the corresponding 2003$-$2011 {\it V} data set, but no significant P$_3$ or P$_4$ was found.  As noted in the overall light curves, the amplitudes in {\it R} are smaller than those in {\it V}.

{\it E. The {\it R$_C$} data set of  the VUO from 1995$-$2011}:
The entire {\it R$_C$} data set was analyzed and yielded values for P$_1$ and P$_2$ that are the same as those found from the entire {\it V} data set and similar to those found for P$_2$ and P$_1$, respectively, in the 2003$-$2011 {\it R$_C$} data set.  The values for P$_3$ and P$_4$ differ from those found for the {\it V} light curve, with P$_4$ barely significant.

{\it F. The {\it V$-$R$_C$} data set of  the VUO from 1995$-$2011}:
The observed {\it V$-$R$_C$} data set from the VUO from 1995 to 2011, which lacks data from 2000 and 2001, was analyzed for periodicity.  Since the mean {\it V$-$R$_C$} values are similar from season to season, no normalization of the data was carried out.  Although the full range in color is small, 0.06 mag, nevertheless clear periods were found.  The values of P$_1$ and P$_2$ are very similar to those of the entire {\it V} and {\it R$_C$} data sets, and P$_3$ is similar to that of the entire {\it V} data set.

We investigated the evidence for a secular period change by comparing the primary period values of the 1994$-$2002 data with those of the 2003$-$2011 data, especially for the {\it V } data, which has the larger amplitude of variation.  From 1994 to 2002, the primary period from the {\it V} light curve is 89.7 days, while for 2003$-$2011, there are two periods of relatively similar amplitudes of 86.7 and 89.5 days.  The {\it R$_C$} light curve from 2003$-$2011 has similar values of P$_1$  and P$_2$.
Thus the results of the period study of these light curves give evidence of multiple periods with some change, but do not give evidence of a secular change in this limited time interval of 18 years of observations.

\begin{figure}\figurenum{13}\epsscale{1.10} 
\plotone{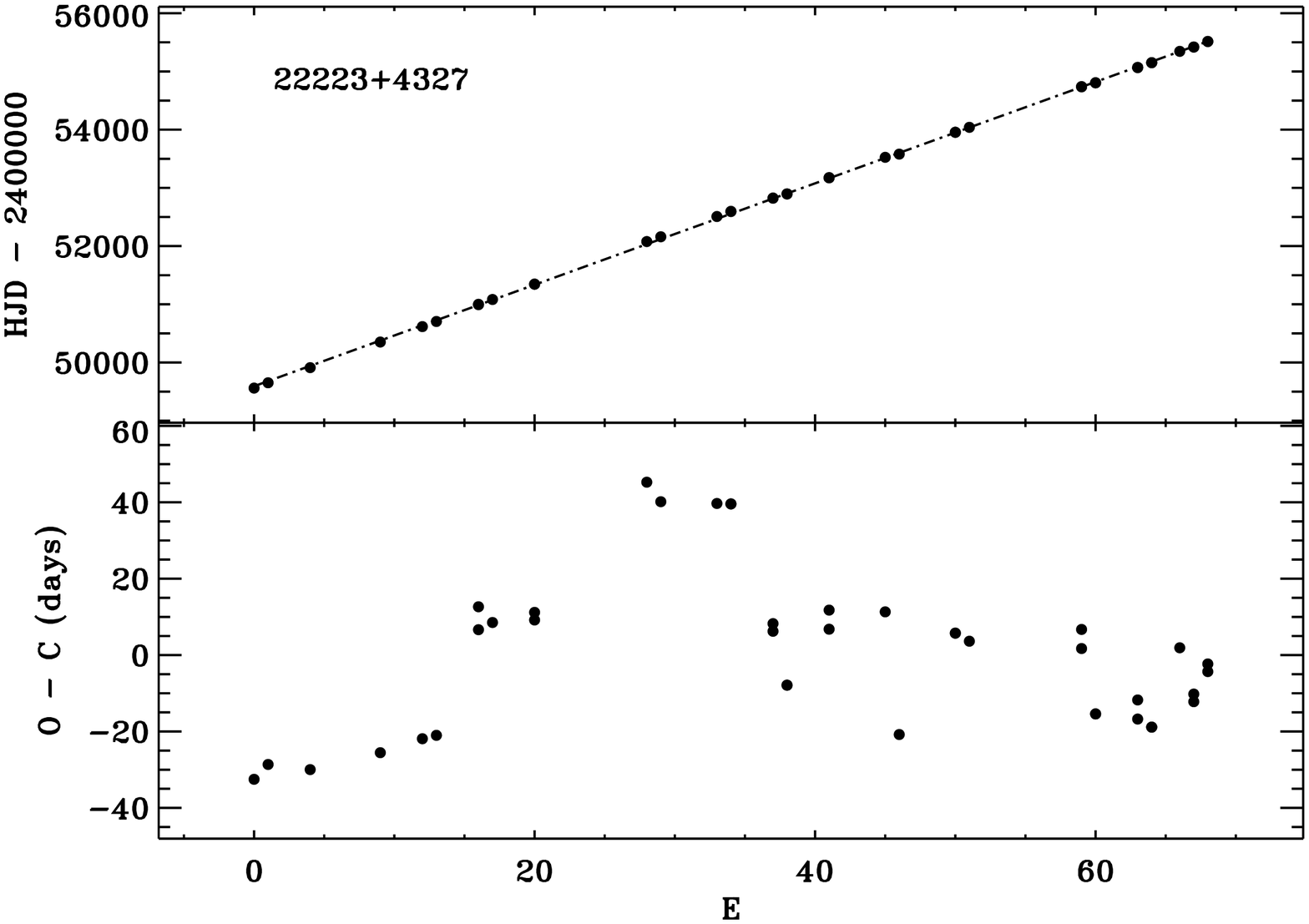}
\caption{Top: Observed time of minimum versus cycle count for the {\it V} and {\it R}$_C$ light curves of IRAS 22223+4327.   
The slope of the line is 87.1 day/cycle.
Bottom: The residuals of the time of minimum about the straight line fit.  
\label{22223_Tmin_E}}
\epsscale{1.0}
\end{figure}

As with IRAS 22272+5435, we also investigated the period change with a plot of cycle number versus Heliocentric Julian Date (HJD) for the minima of the variability.  
This is shown in Figure~\ref{22223_Tmin_E}, based on visual inspection of the VUO {\it V} and {\it R$_C$} light curves.
On this scale it looks linear, with a slope yielding a period of 87.1 days, a value about half way between the two dominant periods found in the periodogram analysis.  
The residuals show a systematic behavior, which could be interpreted as a relatively constant period with a slight increase ({\it E} = 0$-$15), followed by a large increase in period ({\it E} = 15$-$30), then a sudden decrease in period ({\it E} = 35$-$37), and finally a relatively constant period with a slight decrease ({\it E} = 37$-$68).  Note that the full range of the residuals is almost as large as the period.  The pattern does not repeat during the 18 years of our observations. 
As with IRAS 22272+5435, we examined the residuals from the theoretical light curve based on the four periods and corresponding amplitudes, and, as with IRAS 22272+5435, they had a similar range in ({\it O$-$C}) values but not the same pattern.

\placefigure{22223_Tmin_E}

\subsection{Radial Velocity Study}

\begin{figure}\figurenum{14}\epsscale{1.0} 
\plotone{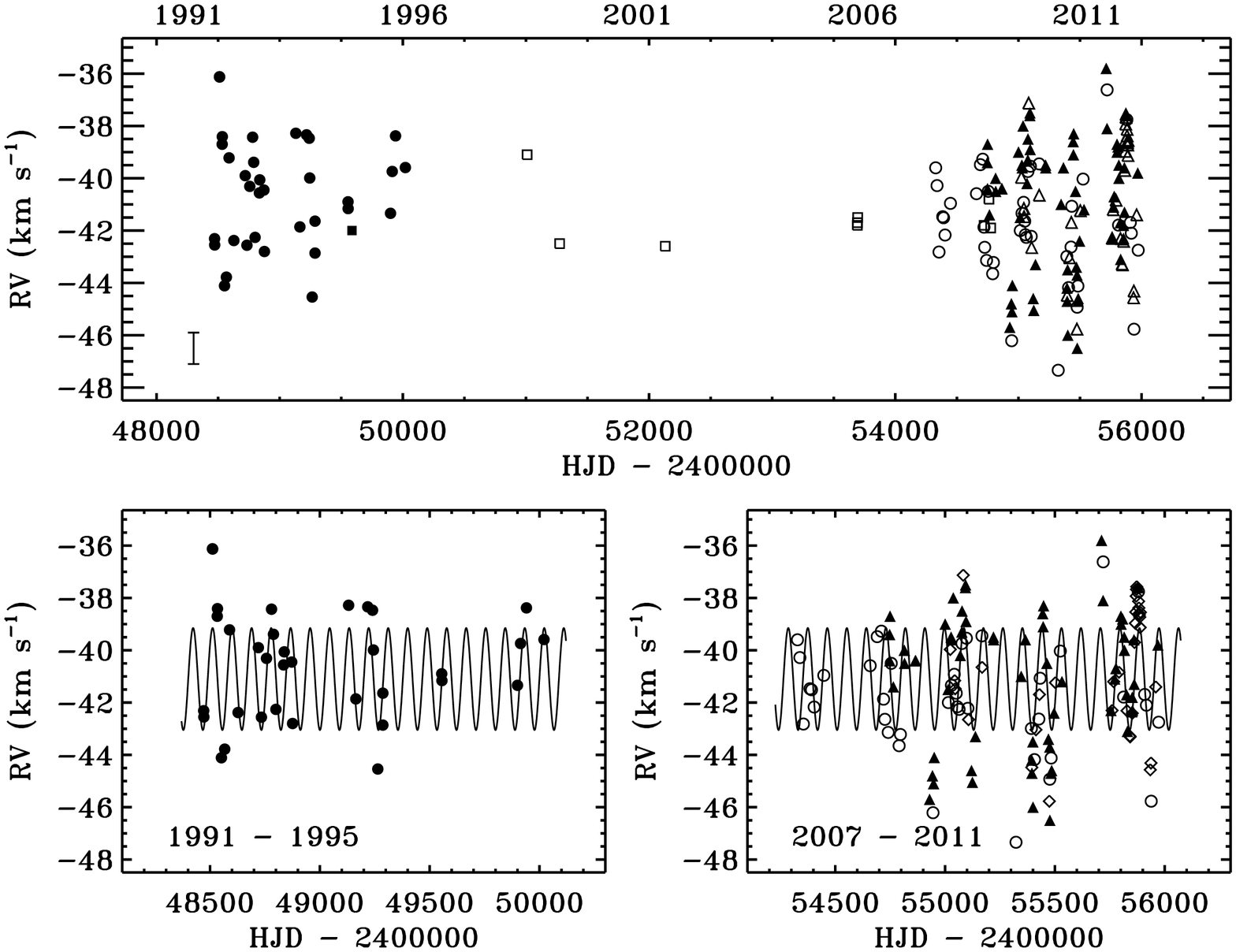}
\caption{Radial velocity observations of IRAS 22223+4327.  The symbols represent the observations made with DAO-RVS (filled circle), DAO-CCD (open circle), CORAVEL (filled triangle), and Hermes (open diamond).  On the lower left is shown a sample error bar for the DAO and CORAVEL data; one for the Hermes data  would be smaller.
Top: The complete radial velocity curve, including the observations of \citet{vanwin00} (filled squares) and \citet{kloch10} (open squares).    
Bottom: Sine curve fits to the velocities based on P = 88.8 d, displayed over shorter intervals of time, 1991$-$1995 and 2007$-$2011.  
\label{22223_rv_P1}}
\epsscale{1.0}
\end{figure}

For the radial velocity study of IRAS 22223+4327, we combined our DAO-RVS observations from 1991$-$1995 with our more recent observations from DAO-CCD (2007$-$2011), CORAVEL (2008$-$2011) and Hermes (2009$-$2011).  
Based on our experience in combining our three recent sets of radial velocity measurements for IRAS 22272+5435, we also inter-compared the measurement for IRAS 22223+4327.
For IRAS 22223+4327, the data are fewer and the results are less certain.  There may be a zero-point shift of approximately +0.5 km~s$^{-1}$ in the CORAVEL and perhaps about half that value in the Hermes measurements as compared with the DAO-CCD measurements, but this is rather uncertain and so no offset was applied.
Thus we combine the data as measured, including the DAO-RVS.
In addition, there exist two published velocities by \citet{vanwin00} from 1994 and nine by \citet{kloch10} from 1998$-$2008, and we have also included them in the velocity analysis.
The combined radial velocity curve is shown in Figure~\ref{22223_rv_P1}.

\placefigure{22223_rv_P1}

We discuss below the results of the analysis of the combined radial velocity measurements and for the data in subsets from 1991$-$1995 and 2007$-$2011.  The results are listed in Table~\ref{22223_vc_per}. 

\placetable{22223_vc_per}

{\it A. The combined radial velocity data set from 1991 to 2011:}  
The data show one dominant period in the data set, P$_1$ = 88.8$\pm$0.1 days. 
This value of P$_1$ is close to the average value of the two main periods determined from the light curves.
A second period of 167.3 days is is found in the velocities but not quite at the level to be regarded as significant.
The value of P$_1$ and the associated amplitude and phase give a reasonably good fit to the observed velocities, as shown in Figure~\ref{22223_rv_P1}.  One can see that at some times the velocity amplitude is not large enough.  Including the second period mentioned above would help somewhat to improve this.  

{\it B. The combined radial velocity data set from 2007 to 2011:} 
This is a reasonably large data set (145 points) and possesses a dominant period of P$_1$ = 88.3$\pm$0.3 days, close to the value found for the entire radial velocity data set.  It gives a good fit to these data.

{\it C. The combined radial velocity data set from 1991 to 1995:} 
This is a small data set of 34 observations spread over five seasons .  
They yield a period of 88.9 days, although at a signal-to-noise ratio that does not quite reach our accepted level of significance.  This gives a good fit to the 1991$-$1995 velocities and the period is in good agreement with that found for the entire data set.

Thus a single dominant period of 88.8 days is found from the radial velocity data, and this value is an approximate (weighted) mean of the two dominant periods found in the light curve data.  The velocity curve can also be fitted reasonably well using the two dominant periods found in the light curve.

\subsection{Contemporaneous Light, Color, \& Velocity Curve Study}

Contemporaneous light, color, and velocity curves exist for IRAS 22223+4327 from 2007 through 2011.  We decided to use for comparison the 2009$-$2011 observations, since for these three seasons the observations are relatively similar.
We did not use the 2007 data since there are only six radial velocity observations, and we did not use the 2008 data since the light, color, and velocity amplitudes were significantly smaller than in the following three years.
For the light and color curves we used only the VUO data; the light and color curves are well defined by these observations and this helps assure a more homogeneous data set. 

\begin{figure}\figurenum{15}\epsscale{1.20}
\plottwo{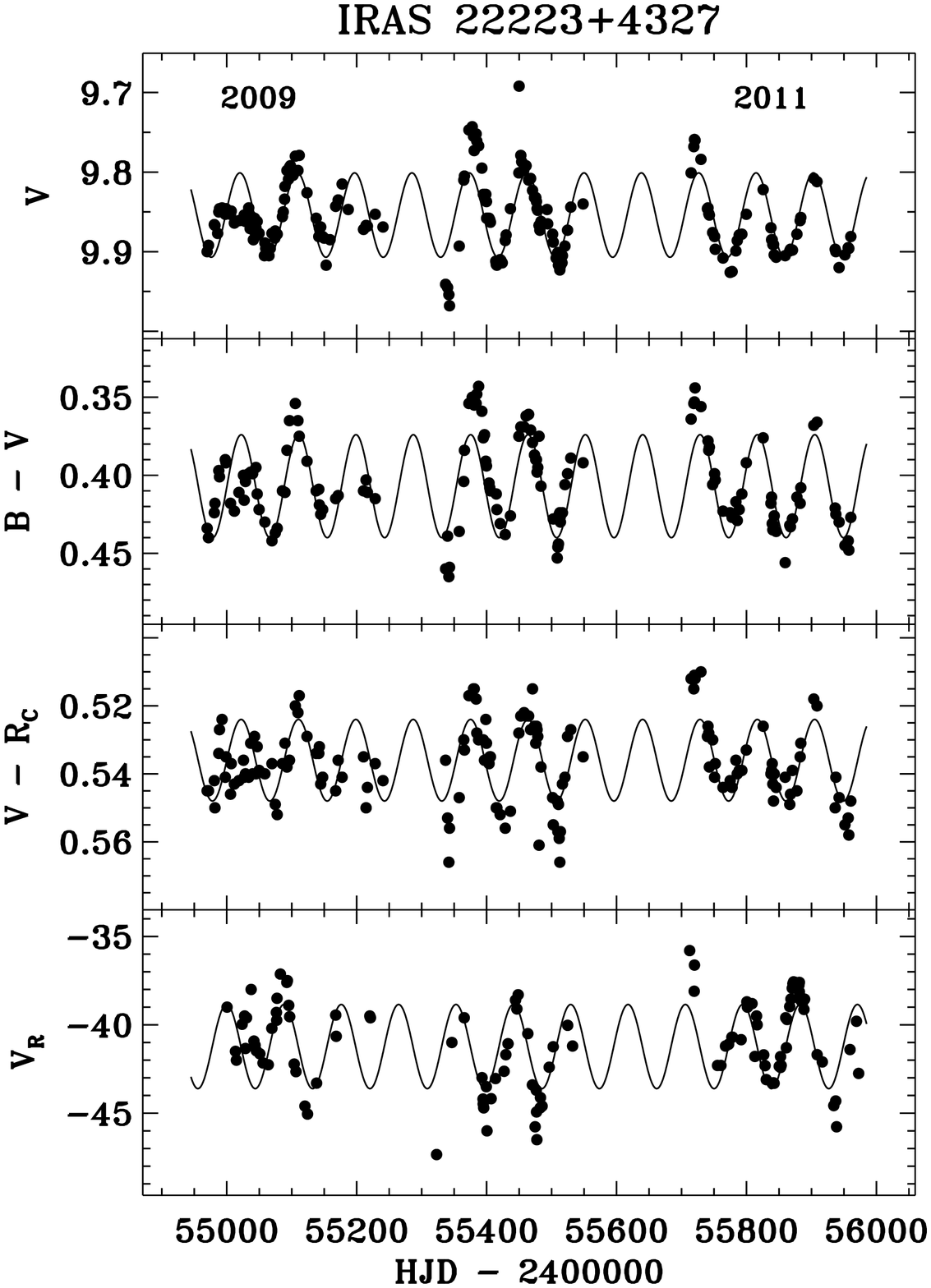}{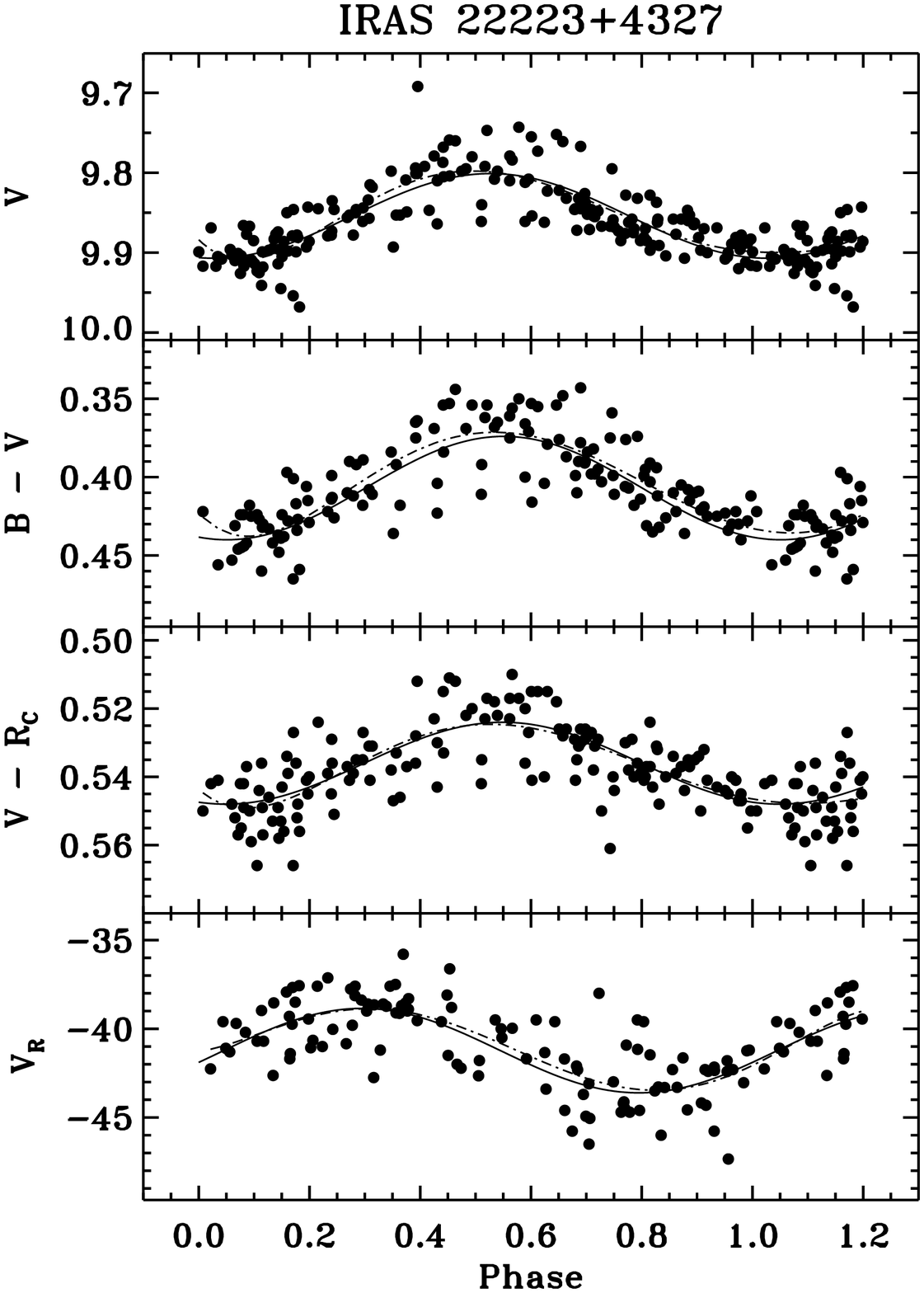}
\caption{(a, left panel) Contemporaneous {\it V}, ({\it B$-$V}), ({\it V$-$R}$_C$), and {\it V$_{\rm R}$} curves of IRAS 22223+4327 from 2009 to 2011, along with sine-curve fits based on P = 88.27 d.
(b, right panel) Phased plots (P = 88.27 d), with polynomial curves (dashed lines) fitted to the observations along with the sine curve fits (solid lines).
\label{22223_2009-11_all}}
\epsscale{1.0}
\end{figure}

These observations are shown in Figure~\ref{22223_2009-11_all}a.
Fitted to these are sine curves based on P = 88.27 days (deduced from the 2009$-$2011 {\it V} light curve) and the arbitrary epoch of 2,448,000.
The curves are fitted with semi-amplitudes and phases, respectively, as follows $-$
{\it V}: 0.053 mag and 0.22, {\it B$-$V}: 0.033 mag and 0.20, {\it V$-$R$_C$}: 0.012 mag and 0.20, and {\it V$_{\rm R}$}: 2.38 km s$^{-1}$ and 0.96.
Thus the light and color curves are very nearly in phase, with the color reaching maximum blueness 0.02 in phase earlier than maximum light, and the radial velocity differs by almost a quarter of a cycle (0.27).
The uncertainties in the phases are $\pm$0.01$-$0.02.
The fit of the {\it V}, {\it B$-$V}, and even the low-amplitude {\it V$-$R$_C$} curves is quite good, and the fit of the {\it V$_{\rm R}$} curve is reasonable. 
Shown in Figure~\ref{22223_2009-11_all}b are the observations phased to this period and epoch.
The observed range in amplitude shows up as vertical scatter in the panels.
Both the sine curve fits and fourth-order polynomial fits are shown for the phased data.  
The polynomials do not differ much from the sine curves.  The slight phase difference between the {\it B$-$V} and {\it V} curves shows up, and the {\it V$_{\rm R}$} curve appears to have a somewhat more gradual decline and somewhat steeper rise than the sine curve.  As with the 2005$-$2007 data of IRAS 22272+5435, the sine curves can be seen to produce good fits to the observations of IRAS 22223+4327 during 2009$-$2011.

\placefigure{22223_2009-11_all}

\section{DISCUSSION}

\subsection{Period Study}

The light curves of both PPNs show evidence of multiple periods that modulate the light curves.  
This appears to not be uncommon, as \citet{kiss07} found evidence for this among several other post-AGB stars.
Four significant periods are found in the light curves of each of these two PPNs, and they can be fit reasonably well with a series of four sine curves.  For the radial velocity curves, which have fewer data points, one to three sine curves give reasonably good fits.  
IRAS 22272+5435 shows excellent agreement between the values of P$_1$ and P$_2$ found for the light curves and velocity curves, with differences less than 1 $\%$.
For IRAS 22223+4327, only one period was determined from the radial velocity curve (88.8 day) and it fell between the values of P$_1$ (90.5 day) and P$_2$ (85.8 day) found from the light curve.

For both of the PPNs, we find from the complete {\it V} light curves a ratio of pulsation periods, P$_2$/P$_1$, of 0.95.  Similar values are found by \citet{ark10,ark11} for two PPNs (0.92, 0.96), by us for several O-rich PPNs \citep{hri13}, 
by \citet {kiss07} for three post-AGB stars (0.90$-$0.94), 
and by \citet{vanwin09} for two post-AGB RV Tau stars (0.91, 0.96). 
It appears that two closely-spaced periods give a good fit to the modulated light curves of these post-AGB objects and may be a characteristic feature of their pulsational variability. 
These are different than the typical ratios between the first overtone and the fundamental modes of 0.71 to 0.73 found for double-mode classical Cepheids in the Milky Way and the Magellanic Clouds \citep{bea97}.  Note that the double-mode Cepheids have periods of only 2 to 6 days and lie near the blue edge of the instability strip.
This difference in pulsation ratio may reflect the differences in the mass or internal structure  between these low-mass post-AGB stars and the more massive Cepheids.

We also investigated the data for evidence of changes in the light curve periods in the $\sim$20 yr of observations.  \citet{hri10} suggested that there may be period decreases of $\sim$2 days over 14 years based on the period$-$temperature relationships that they measured and the model calculations of \citet{stef98} for the evolution of a carbon-rich central star of 0.605 M$_\sun$.  
However, we have not found this.
For IRAS 22272+5435, exactly the same value of the primary period P$_1$ was found in the light curves from 1991$-$1999 and from 2002$-$2011 (although the values of P$_2$ differed).
For IRAS 22223+4327, there appears to be some change in the periods, with the light curve from 1994$-$2002 having dominant periods  of P$_1$ = 89.7 and P$_2$ = 83.3 days, while the light curve from 2003$-$2011 has two periods of more similar strength with P$_1$=86.7 and P$_2$=89.5 days.  
However, this small change is not easy to interpret and does not appear to indicate a shortening of the period.
Also, monotonic period charges are not seen in the plots of the residuals from a linear fit to the observed times of minimum light versus cycle count (Figs. \ref{22272_Tmin_E} and \ref{22223_Tmin_E}).
Such evidence needs to be investigated carefully for more PPNs and likely over longer temporal baselines before one can make a stronger assertion regarding a secular period change.

\subsection{Contemporaneous Light, Color, and Velocity Curves}

The contemporaneous light, color, and velocity curves provide an opportunity to study the pulsation in more detail.  
For IRAS 22272+5435, the observations from 2005$-$2007 appear to be reasonably similar, 
as shown in Figure~\ref{22272_all_2005-07}. 
The {\it V}, {\it V$-$R$_C$}, and radial velocity curves are well represented by sine curves using the parameters listed in Section 5.3.
They show that the light and color curves have the same phasing and that the radial velocity curve differs in phase by $-$0.26, almost exactly a quarter of a cycle.
This indicates that the star is brightest when hottest and smallest.  
Polynomial fits to the data are also shown.  
For the radial velocity curves, the two different fits are very similar.
For the light and color curves, the polynomial fits have similar shapes to each other and both differ slightly from the sine fits.  The polynomial fits reach maximum a little earlier and minimum a little later in phase, resulting in a slightly slower decline to minimum and a steeper rise to maximum.
Contemporaneous observations for IRAS 22272+5435 also exist from 2008$-$2011, and we have examined the more similar years 2008$-$2009.  While these data are not so well fitted by sine curves, they do indicate that on average the light and color are nearly in phase and the radial velocity curve again about a quarter of a cycle ($-$0.22) different.
A similar comparison can be made for the contemporaneous observations of IRAS 22223+4327, and we have done so using the data from 2009$-$2011.
One again finds that the light and color curves are nearly in phase (differing by $-$0.02) and the radial velocity differs by about a quarter of a cycle ($-$0.27).
Again, comparison with polynomial fits show that sine curves are a good representations of the variations.
Overall, given the intrinsic changes in the variability from season to season and even over a single season, to within the uncertainty one can state that the curves are all well represented by sine curves, with the light and color curves in phase and the radial velocity curve differing in phase by a quarter of a cycle ($-$0.25).

\subsection{Comparison with Other Evolved Pulsating Stars}

These observed curves are quite different from those of classical Cepheids pulsators in the following ways:
(1) in their shapes $-$ in Cepheids the light and color curves show a steeper rise and a more gradual decline, with the velocity curves showing the opposite; 
(2) in the phasing between the light and color curves $-$ in Cepheids the color minimum is later than the light minimum;
(3) in the phasing of the light and velocity curve $-$ in Cepheids they are the mirror images of each other, differing in phase by 0.50.  Thus in Cepheids the photosphere is moving toward the observer at maximum speed and at approximately average size when it is at light maximum, while in these PPNs the photosphere is temporarily stationary at a minimum size at light maximum.  
In other words, these two PPNs do not display the phase lag seen in classical Cepheids, which is caused by the delaying action of the hydrogen partial ionization zone on the energy flux \citep{ost07}.

RV Tauri variables have more similar masses and might be a better comparison.  They have mid-F$-$K spectral types and spectra of luminosity classes I $-$II.  Their light curves are characterized by alternating deeper and shallower minima, with periods between deeper minima of 50$-$150 day \citep{wah93}.  Strong Balmer-line emission and metallic-line doubling is observed and attributed to shocks.  
A comparison of the light, color, and velocity curves is complicated by the variation in these curves from cycle to cycle and by the lack of contemporaneous light and velocity curves.
\citet{mcl41}, in an old study, attempted to define from observations a generalized light and velocity curve for an ``ideal RV Tauri variable.''  He concluded that the velocity maximum occurs a few days after the deeper light  minimum.  In his associated figure (Fig. 6), it appears that the deeper minimum occurs when the star is close to, but slightly after, largest size and that the brighter maximum occurs when the star is close to, but slightly before, minimum size.
This is more similar to what we see in these two PPNs.
However, there is a significant difference in that the light and velocity curves of RV Tauris display alternating deeper and shallower light and velocity minima that are not characteristic of the PPN light and velocity curves.
\citet{poll96,poll97} carried out a comprehensive photometric and spectroscopic study of 11 RV Tauri variables over two to four seasons.  These objects showed a much larger range of variation in brightness (0.8 to 3.0 mag in {\it V}) and radial velocity (15 to 50 km~s$^{-1}$) than is seen in these two or other PPNs \citep{hri10,hri11}.  They did not have good enough contemporaneous coverage of light and velocity to do an accurate comparison of the phasing, so unfortunately we cannot compare the relative phasing of the RV Tauris with that found for these two PPNs.  The color curves do show that for RV Tauris the light is bluest on the rise to maximum light, preceding the light peak by $\sim$0.1 in phase.  Our PPN data may show a slight tendency for to be bluer before the light maximum, but only by $\sim$0.02 in phase, which is at the level of our uncertainty.

\subsection{Determination of Radii and Luminosities}

One can use the observed radial velocity curve to calculate the change in radius of the pulsating star.
Since the radial velocity curve is well represented by a sine curve, we have simply used this to calculate the change in radius throughout the pulsation cycle.
Integrating the radial velocity curve for IRAS 22272+5435 with its semi-amplitude of 3.02 km s$^{-1}$ over the period of 131.9 d yields a semi-amplitude for the radius curve of 5.48 $\times$ 10$^6$ km, or 7.9 R$_\sun$.  
To convert the radial velocity to pulsational velocity requires one to correct for geometric projection, atmospheric effects, and other complications \citep{bar09}.  Using the Cepheid equation of \citet{nar09} based on cross-correlation velocities, {\it p} = 1.31($\pm$0.06) $-$ 0.08($\pm$0.05){\it logP}, gives {\it p} = 1.15.  Assuming this value for {\it p} yields a total change in radius of 1.26 $\times$ 10$^7$ km, or 18.1 R$_\sun$.  
This is a very large value. 
The brightness only changes by 0.34 mag ({\it V}) and the color varies by 0.076 in ({\it V$-$R$_C$}) during this time.

For IRAS 22223+4327, the radial velocity semi-amplitude of 2.38 km s$^{-1}$ and period of 88.3 d result in a semi-amplitude of the radius curve of 2.89 $\times$ 10$^6$ km, or 4.2 R$_\sun$, and with the above projection factor leads to a total radius change of 9.6 R$_\sun$.  This is coupled with a brightness change of 0.106 mag ({\it V}) and color changes of 0.066 mag in ({\it B$-$V}) and 0.024 mag in ({\it V$-$R$_C$}).
 
We attempted to use these changes in size, along with changes in brightness and temperature (from color) to determine the linear size and luminosity of the stars.  
Assuming that the temperature determined in the abundance analyses listed in Table~\ref{object_list} is an average temperature 
and assuming a color-temperature relationship for PPNs similar to that of supergiants, which is a reasonable approximation given the log {\it g} values for these two PPNs of 0.5 and 1.0, 
one finds that the color range of $\Delta$({\it V$-$R$_C$}) = 0.076 for IRAS 22272+5435 
(T$_{\rm eff}$=5750 K) translates to a temperature variation of $\Delta$T$_{\rm eff}$ = 476 K  and a variation in bolometric correction of $\Delta$({\it BC}) = $-$0.057 \citep{cox00}.  
For IRAS 22223+4327 (T$_{\rm eff}$=6500 K), one finds that the color range of $\Delta$({\it V$-$R$_C$}) = 0.024 translates to $\Delta$T$_{\rm eff}$ = 170 K and $\Delta$({\it BC}) = $-$0.011.   
Knowing the difference in {\it V} brightness from maximum to minimum and the difference in bolometric correction at these two phases based on the temperature change, one can find the luminosity ratio between maximum and minimum brightness.  
Given that the stars are brightest and hottest when smallest, and faintest and coolest when largest, and assuming that the T$_{\rm eff}$ is the average temperature
and that the stars behave like  black bodies, we can then find the ratio of minimum to maximum radius.  Knowing the ratio and the difference allows us to then calculate the radius of each of these PPNs.
For IRAS 22272+5435, this leads to a ratio of luminosities of 0.773, a ratio of radii of 1.038, and an average radius of 555 R$_\sun$.  This radius value is much larger than that for a post-AGB star of this temperature, and leads to a luminosity of 3.0$\times$10$^5$ L$_\sun$.  
Such a luminosity is that of a yellow supergiant or hypergiant. It is much larger than that of a post-AGB star, which from theoretical models fit to the luminosities of PN is expected to be in the range of 4-10$\times$10$^3$ L$_\sun$ \citep{blo95}, and which is determined for LMC carbon-rich PPNs to have a value of 4-10$\times$10$^3$ L$_\sun$ \citep{volk11}.
Assuming a typical of {\it L} = 8$\times$10$^3$ L$_\sun$ leads to a radius of 90 R$_\sun$ and would require a larger variation in brightness and color (temperature) or a smaller range in radius of 3.4 R$_\sun$, one-fifth of the value determined from the radial velocity curve.
A similar discrepancy occurs for the parameters of IRAS 22223+4327.

On the other hand, one gets somewhat smaller than expected values for the luminosity and radius when using the spectroscopically determined T$_{\rm eff}$ and log~{\it g} values.  These lead to values of 5100 L$_\sun$ and 72 R$_\sun$ for IRAS 22272+5435 and 2600 L$_\sun$ and 41 R$_\sun$ for IRAS 22223+4327 for M = 0.60 M$_\sun$, with values of luminosity and radius that are larger by 1.33 and 1.15, respectively, for M = 0.80 M$_\sun$.
Clearly there are unresolved complications in both the pulsational light and velocity curves, and even in the spectroscopically-determined physical parameters, that are not presently understood.

The previous spectroscopic study by \citet{zacs09} had demonstrated the presence of shocks in the pulsating atmosphere of IRAS 22272+5435.  This was revealed by line profile variations and the doubling of low-excitation lines.  Earlier observations of changes between absorption and emission in near-infrared CO lines in IRAS 22272+5435 and 22223+4327 can be attributed to sporadic mass loss or pulsation causing shocks \citep{hri94}.  So it is clear that we are not dealing with static atmospheres in thermal equilibrium, and more complex models are required to understand their behavior and properties.  
Similar or perhaps even greater complexity and evidence for shocks have previously been found in the related C-rich PPN IRAS 07134+1005 \citep[HD~56126;][]{bar00}.  This object is hotter \citep[T$_{\rm eff}$=7250 K;][]{vanwin00}, with a spectral type of $\sim$F5~I and a dominant period of 35$-$40 d \citep{hri10,bar00}.  It is likely a more evolved analog of the two PPNs in the present study.
 
We had anticipated trying to use the Baade$-$Wesselink Method to find the radius of each star. This method uses the difference in luminosity in the pulsation curve at two different phases when the color values, and thus the temperatures, are the same to find the ratio of radii.  But since the color and luminosity curves are in phase and similar in shape (approximately sinusoidal), the luminosities are the same when the colors are the same and thus the radii are the same at those phases and no ratio can be determined.  However, given the unresolved complexities that are manifest in the unrealistically large values derived from the amplitudes of the light and velocity curves above, and even the low values derived from the spectroscopic analysis, the prognosis for obtaining reliable radii from the Baade$-$ Wesselink Method would be low.\\

\subsection{Comparison with Pulsational Models}

These observational results can be compared with pulsational models of post-AGB stars of similar temperatures.

\citet{fok01} computed non-linear, radiative models for post-AGB stars in the temperature range 5600$-$6000 K and mass of 0.6 and 0.8 M$_\sun$.  Several patterns emerged: the photometric amplitude increased as temperature decreased, the chaotic components decreased as the luminosity decreased and as the temperature increased, the power in the lower modes increased as the temperature decreased, and the photometric amplitude increased as the mass of the star decreased.  
They sought to apply these models to the shorter period ($\sim$37 days) PPN IRAS 07134+1005 (HD 56126), which is a carbon-rich F star.  They had mixed success.
Their models agreed with the period and the amplitude of the radial velocity curve, but produced a larger photometric amplitude than observed in this object and had a mass that was too large (0.8 M$_\sun$) to be consistent with the luminosity (6000$-$7000 L$_{\sun}$) that they determined.  Also, the temperature of the models was much lower than the spectroscopically determined value of the effective temperature, 7250 K \citep{vanwin00}.
Their models are of the appropriate temperature for IRAS 22272+5435, and give a reasonable fit to both the photometric and radial velocity amplitudes.  However, their calculated periods ($<$40 days) are much shorter than the observed period of $\sim$130 days.

Recently \citet{aik10} published the results of a study of pulsation in post-AGB stars of the following range in temperature and gravity: log~{\it T$_{\rm eff}$} = 3.7$-$3.9 (5000$-$8000 K) and log~{\it g} = 0.0$-$1.8, for masses of 0.6 and 0.8 M$_{\sun}$.  The models assumed radial pulsation, radiative energy transfer, and ignored convection.
He ran linear, radial pulsation models and studied the stability from the fundamental up to the fifth overtone mode.   
For the temperature and gravity of our two stars, the appropriate models generally were found to be stable in the fundamental and first overtone modes and unstable in the higher order (2$-$5) overtone modes.
For the closest model for IRAS 22272+5435 (for which log~{\it T$_{\rm eff}$} = 3.76), the log~{\it T$_{\rm eff}$} = 3.75 model produced a fundamental period of 62 d (68 d) for M = 0.6 M$_{\sun}$ (0.8 M$_{\sun}$), with a first overtone period of 26 d (35 d).
For IRAS 22223+4327 (log~{\it T$_{\rm eff}$} = 3.81), the log~{\it T$_{\rm eff}$} = 3.80 model produced a fundamental period of 26 d (25 d) for M = 0.6 M$_{\sun}$ (0.8 M$_{\sun}$), with a first  
overtone period of 14 d (14 d).
These computed values of the fundamental period are on the order of one-half to one-third of the observed periods and the overtones are correspondingly even shorter.
Non-linear simulations were then run to produce light curves for the unstable modes. 
These photometric light curves have short periods and low amplitudes, 14 days and 0.20$-$0.25 mag peak-to-peak
for the best-fit model for IRAS 22272+5435 and 
4 days and 0.06 mag peak-to-peak for the best-fit model for IRAS 22223+4327.

Thus these published models are in poor agreement with the observed pulsation periods 
of these two cool post-AGB stars.  Clearly additional modeling efforts, ideally including convection, are needed to find agreement with the observations.

\section{SUMMARY AND CONCLUSIONS}

In this paper, we reported on an intensive photometric and radial velocity study of two PPNs, IRAS 22223+4327 and 22272+5435.  Yearly multicolor light curves from 1994 to 2011 are combined with radial velocities from the intervals 1991 to 1995 and 2005 or 2007 to 2011. 
The primary results of this study are listed below.

1. The light curves display evidence of multiple periods that modulate the light curves.
Each contained four significant periods and the radial velocity curves, which are not based on as extensive data sets, contained one period (IRAS 22223+4327) or three periods (IRAS 22272+5435).  The periods determined from the radial velocity measurement agree with those determined from the light curves.

2. The dominant primary periods of the two PPNs are 90.5 days for IRAS 22223+4327 and 131.9 days for IRAS 22272+5435.  The ratio of secondary to primary periods, based on the light curves, is 0.95 for each of the two.  This value has also been found for several other post-AGB stars and may be a characteristic of post-AGB pulsational variability.  It is  much different from the ratio of $\sim$0.72 found for Cepheid variables.

3. Comparing contemporary light, color, and radial velocity curves for several years reveals that the light and color curves are in phase, with the objects cooler when fainter and no significant phase lag between the two.  The radial velocity curves are 0.25 {\it P} out of phase with the light curve, with the objects smallest when hottest.  The semi-amplitudes $\Delta${\it V} and $\Delta${\it V$_r$} are small, 0.11 mag and 2.4 km~s$^{-1}$ for IRAS 22223+4327 and 0.34 mag and 3.0 km~s$^{-1}$ for IRAS 22272+5435 for the contemporaneous curves, with shapes that are close to sinusoidal.  
These are in contrast to Cepheid variables, in which there is a slight phase lag between the color and the light curves and in which the radial velocity curves differ by 0.5 {\it P} from the light curves, with the Cepheids having average size and expanding when the stars are at maximum brightness. 

4. An attempt to use the contemporaneous curves to determine the luminosity and radii of the PPNs led to values that are much too large for PPNs.  The source of these discrepancies requires further investigation.

5. These values for the observed periods are much longer than those calculated by nonlinear pulsation models and the observed amplitudes are generally somewhat larger.  These point to the need for new models, at least in some cases including convection, to try to understand the pulsational nature of these post-AGB stars that are in transition from the AGB to the planetary nebula phases.  Such models have the potential to allow the determination of the mass and luminosity of these objects, which presently have no direct means of being determined.

This combination of light, color, and radial velocity observations, especially when they are made contemporaneously, allows one to better constrain the pulsational properties of these stars.  The fact that the light curves are complicated, with changing amplitudes and changing spacing between minima, makes it necessary that such long temporal data sets be obtained to sort out the component periods and amplitudes.  This study is a first step in that direction, and we are in the process of carrying out similar studies of other bright PPNs.

\acknowledgments  
We gratefully acknowledge helpful conversations with P. Lenz regarding Period04
and with D. Turner about variable stars in general. 
The comments of the anonymous referee were appreciated.
We thank R. D. McClure and C. D. Scarfe for making some observations for us, 
K. L. Wefel who assisted in the reduction of the early DAO radial velocity data, 
and the many Valparaiso University undergraduate students,
most recently Kristie Shaw, Wesley Cheek, Rachael Jensema, Ryan McGuire, Christopher Miko, Zachary Nault, Joel Rogers, Samuel Schaub, and Christopher Wagner, 
who carried out the photometric observations over these past 18 years.
We acknowledge support for this collaboration from the EU FP7-PEOPLE-2010-IRSES program in the framework of project POSTAGBinGALAXIES (Grant Agreement No.269193).
BJH acknowledges the support of a University Research Professorship and a 
sabbatical leave from Valparaiso University and the hospitality of the 
Dominion Astrophysical Observatory during the initial stages of this research project.
BJH also acknowledges support from the National Science Foundation
(AST 9018032, 9315107, 9900846, 0407087, 1009974), NASA through the JOVE program, 
and the Indiana Space Grant Consortium.
LZ acknowledges support from the Latvian Council of Science under grant No. 09.6190, and
JS and LZ acknowledge support from the Research Council of Lithuania under the grant MIP-085/2012.  
This research has made use of the SIMBAD database, operated at CDS, Strasbourg,
France, and NASA's Astrophysical Data System.

\clearpage

\tablenum{1}
\begin{deluxetable}{crrrrrrrcl}
\tablecaption{Program Objects\label{object_list}} 
\tablewidth{0pt} \tablehead{
\colhead{IRAS ID}&\colhead{{\it V}\tablenotemark{a}}&\colhead{({\it V$-$R}$_C$)\tablenotemark{a,b}}&&\colhead{Sp.T.}&\colhead{T$_{\rm eff}$}&\colhead{log {\it g}}&\colhead{[Fe/H]}
&\colhead{C/O}&\colhead{Ref.\tablenotemark{c}}\\
\colhead{}&\colhead{(mag)}&\colhead{(mag)}&&\colhead{}&\colhead{(K)}
&\colhead{}&\colhead{}&\colhead{}&\colhead{}} \startdata
22223+4327 & 9.8 & 0.54 && G0 Ia & 6500 & 1.0 & $-$0.3 & 1.2 &1 \\
22272+5435 & 8.6 & 1.03 && G5 Ia & 5750 & 0.5 & $-$0.8 & 1.6 & 2 \\
\enddata
\tablenotetext{a}{Variable. }
\tablenotetext{b}{Includes circumstellar and interstellar reddening. }
\tablenotetext{c}{References for the spectroscopic analyses: (1) \citet{vanwin00}, (2) \citet{red02}. }
\end{deluxetable}

%\clearpage

\tablenum{2}
\begin{deluxetable}{llrrrc}
\tablecaption{Standard Magnitudes of Program and Comparison
Stars\tablenotemark{a} \label{phot_std}} 
\tablewidth{0pt} \tablehead{
\colhead{Object} &\colhead{GSC ID} &\colhead{{\it V}} &\colhead{{\it B$-$V}} &\colhead{{\it V$-$R}$_C$} &\colhead{Run\tablenotemark{b}}}
\startdata
IRAS~22223+4327 & 03212-00676 &   9.88 & 1.00  & 0.54 & 1 \\
			      &                    &   9.85 & 0.93  & 0.54 & 2 \\
C$_1$                    & 03212-00672 & 11.08 & 0.54 & 0.31 & \nodata \\
C$_2$                    & 03212-00561 & 11.88 & 0.33 & 0.18 & \nodata \\
C$_3$                    & 03212-00499 & 12.18 & 1.16 & 0.62 & \nodata \\
\\
IRAS~22272+5435  & 03987-01344 &  8.56 & 2.07 & 1.01 & 1 \\
			      &		      &  8.55 & 1.94 & 1.00 & 2 \\
C$_1$                     & 03987-02320 &11.63 & 0.58 & 0.33 & \nodata \\
C$_2$                     & 03987-00944  &11.38 & 0.34 & 0.17 & \nodata \\
C$_3$                     & 03987-00512 &11.16 & 1.11 & 0.60 & \nodata \\
\enddata
\tablenotetext{a}{Uncertainties in the observations are as follows $-$ {\it V}: $\pm$0.01, 
{\it B$-$V}: $\pm$0.025, and {\it V$-$R}$_C$: $\pm$0.015 mag. }
\tablenotetext{b}{Observations made at the VUO on (1) 20 May 2009 and (2) 26 June 2012 (UT), and are averaged together for the comparison stars. }
\end{deluxetable}

\clearpage

\tablenum{3}
\begin{deluxetable}{rrrrcrrrr}
\tablecaption{Differential Standard Magnitudes for IRAS 22223+4327\tablenotemark{a} \label{new_phot_22223}}
\tablewidth{0pt} \tablehead{ \colhead{HJD$-$2,400,000\tablenotemark{b}}
&\colhead{$\Delta${\it B}}  &\colhead{$\Delta${\it V}} &\colhead{$\Delta${\it R}$_C$} & &
\colhead{HJD$-$2,400,000\tablenotemark{b}}
&\colhead{$\Delta${\it B}}  &\colhead{$\Delta${\it V}} &\colhead{$\Delta${\it R}$_C$}}
\startdata
54654.7279  &  \nodata &  \nodata & -1.430  && 55142.5341 & -0.792 &  -1.215 & -1.427\\
54662.7959  & -0.777 & -1.204 & -1.426  && 55144.5962 & -0.787 &  -1.216 & -1.439\\
54663.8319  & -0.792 & -1.208 & -1.431  && 55147.6041 & -0.777 &  -1.204 & -1.425\\
54664.7696  & -0.788 & -1.207 & -1.427  && 55167.6577 & -0.823 &  -1.242 & -1.467\\
54665.6989  & -0.787 & -1.193 & -1.411  && 55171.5303 & -0.834 &  -1.250 & -1.466\\
54670.7185  & -0.795 & -1.210 & -1.432  && 55177.5625 &  \nodata  &  -1.270 & -1.491\\
54672.7058  & -0.781 & -1.209 & -1.432  && 55210.5381 & -0.799 &  -1.213 & -1.428\\
54689.7864  & -0.817 & -1.247 & -1.470  && 55214.4813 & -0.810 &  -1.218 & -1.448\\
54694.7635  & -0.848 & -1.248 & -1.467  && 55216.5072 & -0.802 &  -1.217 & -1.441\\
54696.7320  & -0.859 & -1.256 & -1.462  && 55228.5020 & -0.813 &  -1.232 & -1.449\\
\enddata
\tablenotetext{a}{Table~\ref{new_phot_22223} is published in its entirety in the 
electronic edition of the Astrophysical Journal.  A portion of Table~\ref{new_phot_22223} 
is shown here for guidance regarding form and content.}
\tablenotetext{b}{The time represents the mid-time of the {\it V} observations.  The times for the {\it B} and {\it R}$_C$ observations differ from this by approximately +0.0034 and $-$0.0020 day, respectively.}
\end{deluxetable}

%\clearpage

\tablenum{4}
\begin{deluxetable}{rrrrcrrrr}
\tablecaption{Differential Standard Magnitudes for IRAS 22272+5435\tablenotemark{a} \label{new_phot_22272}}
\tablewidth{0pt} \tablehead{ \colhead{HJD$-$2,400,000\tablenotemark{b}}
&\colhead{$\Delta${\it B}}  &\colhead{$\Delta${\it V}} &\colhead{$\Delta${\it R}$_C$} & &
\colhead{HJD$-$2,400,000\tablenotemark{b}}
&\colhead{$\Delta${\it B}}  &\colhead{$\Delta${\it V}} &\colhead{$\Delta${\it R}$_C$}}
\startdata
54654.7077 &  \nodata  &  \nodata   & -3.784 && 55138.6160 & -1.703 & -3.122 & -3.780\\
54662.7986 & -1.476 & -2.982 & -3.668 && 55141.6836 & -1.687 & -3.116 & -3.799\\
54663.8366 & -1.497 & -2.973 & -3.657 && 55142.5434 & -1.664 & -3.098 & -3.805\\
54665.7128 & -1.489 & -2.971 & -3.657 && 55144.6043 & -1.653 & -3.091 & -3.755\\
54670.7485 & -1.405 & -2.935 & -3.635 && 55147.6091 & -1.623 & -3.058 & -3.742\\
54672.7231 & -1.384 & -2.909 & -3.595 && 55167.7125 & -1.462 & -2.938 & -3.606\\
54689.7924 & -1.204 & -2.791 & -3.511 && 55171.5371 & -1.436 & -2.902 & -3.600\\
54694.7666 & -1.219 & -2.775 & -3.484 && 55182.5854 & -1.430 & -2.891 & -3.575\\
54696.7369 & -1.246 & -2.780 & -3.504 && 55210.5726 & -1.423 & -2.909 & -3.569\\
54697.8296 & -1.214 & -2.757 & -3.473 && 55214.4919 & -1.399 & -2.891 & -3.584\\
\enddata
\tablenotetext{a}{Table~\ref{new_phot_22272} is published in its entirety in the 
electronic edition of the Astrophysical Journal.  A portion of Table~\ref{new_phot_22272} 
is shown here for guidance regarding form and content.}
\tablenotetext{b}{The time represents the mid-time of the {\it V} observations.  The times for the {\it B} and {\it R}$_C$ observations differ from this by approximately +0.0025 and $-$0.0010 day, respectively.}
\end{deluxetable}

\clearpage

\tablenum{5}
\begin{deluxetable}{cccccccc}
\tablecaption{Radial Velocity Observations of
IRAS 22223+4327\label{tab_22223RV}}
\tablewidth{0pt} \tablehead{ \colhead{HJD$-$2,400,000}
&\colhead{{\it V}$_ r$} 
&&\colhead{HJD$-$2,400,000} &\colhead{{\it V}$_ r$} 
&&\colhead{HJD$-$2,400,000}
&\colhead{{\it V}$_ r$} \\
\colhead{}&\colhead{(km s$^{-1}$)}&&\colhead{}&\colhead{(km s$^{-1}$)}&&\colhead{}&\colhead{(km s$^{-1}$)}  }
\startdata
&&& DAO-RVS &&&& \\
\tableline
48470.9777 & -42.31 && 48756.9547 & -40.31 && 49243.9289 & -39.99 \\
48471.8874 & -42.55 && 48779.9297 & -38.43 && 49263.8801 & -44.54 \\
48510.8859 & -36.12 && 48788.8747 & -39.39 && 49286.8661 & -41.64 \\
48511.8290 & -36.13 && 48799.9333 & -42.26 && 49286.8995 & -42.86 \\
48532.7969 & -38.70 && 48834.8871 & -40.56 && 49554.9029 & -40.90 \\
48533.8200 & -38.41 && 48837.8878 & -40.06 && 49555.9399 & -41.16 \\
48551.7917 & -44.11 && 48871.8468 & -40.45 && 49899.9354 & -41.34 \\
48566.7475 & -43.78 && 48875.8244 & -42.80 && 49913.9602 & -39.74 \\
48588.7468 & -39.22 && 49130.9690 & -38.28 && 49940.8221 & -38.38 \\
48627.6548 & -42.38 && 49163.9382 & -41.86 && 50020.6870 & -39.59 \\
48719.9993 & -39.90 && 49217.9829 & -38.34 && \nodata & \nodata \\
48733.9357 & -42.56 && 49239.8757 & -38.47 && \nodata & \nodata \\
\tableline
&&& DAO-CCD &&&& \\
\tableline
54327.8994 & -39.60 && 54790.6756 & -43.65 && 55392.9142 & -42.99 \\
54338.8103 & -40.28 && 54796.6355 & -43.22 && 55406.9462 & -44.18 \\
54354.8812 & -42.82 && 54944.9767 & -46.21 && 55426.9406 & -42.63 \\
54384.7518 & -41.47 && 55014.8890 & -42.00 && 55432.9379 & -41.07 \\
54391.7458 & -41.51 && 55028.8741 & -41.34 && 55476.8042 & -44.93 \\
54403.7580 & -42.17 && 55041.8754 & -40.92 && 55482.8715 & -44.12 \\
54447.7227 & -40.96 && 55050.9065 & -41.64 && 55524.7324 & -40.03 \\
54658.9294 & -40.59 && 55055.8915 & -42.16 && 55719.9204 & -36.62 \\
54690.8693 & -39.48 && 55063.8680 & -42.26 && 55812.9217 & -41.79 \\
54709.8763 & -39.27 && 55076.8689 & -39.74 && 55880.6686 & -37.76 \\
54719.9476 & -41.87 && 55096.8159 & -39.54 && 55908.7026 & -41.69 \\
54725.9390 & -42.64 && 55103.8066 & -42.22 && 55916.6871 & -42.10 \\
54740.9688 & -43.14 && 55167.7556 & -39.45 && 55938.6286 & -45.77 \\
54754.7945 & -40.50 && 55322.9618 & -47.34 && 55972.6245 & -42.75 \\
\tableline
&&& CORAVEL &&&& \\
\tableline
54747.3690 & -39.40 && 55095.4120 & -38.90 && 55532.3040 & -41.20 \\
54748.3000 & -40.40 && 55120.3360 & -44.60 && 55712.5140 & -35.80 \\
54749.4260 & -38.70 && 55124.3340 & -45.05 && 55719.4970 & -38.10 \\
54764.4260 & -41.40 && 55138.2500 & -43.30 && 55755.4370 & -42.30 \\
54814.3500 & -40.00 && 55220.1830 & -39.50 && 55772.4990 & -41.10 \\
54815.2570 & -40.50 && 55221.1810 & -39.60 && 55777.4870 & -40.70 \\
54864.1970 & -40.40 && 55346.5130 & -41.00 && 55778.5180 & -40.70 \\
54866.1960 & -40.40 && 55365.4980 & -39.60 && 55800.5230 & -38.70 \\
54929.5890 & -45.70 && 55394.4950 & -44.20 && 55801.4590 & -39.00 \\
54942.5640 & -44.80 && 55395.5130 & -44.70 && 55808.4860 & -38.80 \\
54947.5480 & -45.10 && 55399.5210 & -43.50 && 55815.4510 & -39.50 \\
54950.5560 & -44.10 && 55400.5260 & -46.00 && 55816.3910 & -40.00 \\
55000.4800 & -39.00 && 55444.4920 & -38.60 && 55826.4830 & -41.70 \\
55013.4660 & -41.50 && 55446.5340 & -39.10 && 55830.3770 & -43.10 \\
55027.5050 & -39.50 && 55448.5400 & -38.30 && 55850.4120 & -42.35 \\
55030.4750 & -39.60 && 55463.4160 & -40.50 && 55852.4220 & -41.80 \\
55037.5250 & -38.00 && 55470.4480 & -43.40 && 55853.3290 & -42.30 \\
55069.4300 & -40.20 && 55476.4080 & -43.70 && 55860.3200 & -39.60 \\
55076.4290 & -39.30 && 55477.3080 & -46.50 && 55861.3980 & -41.30 \\
55077.3640 & -38.50 && 55482.4150 & -44.70 && 55875.3610 & -37.60 \\
55092.4850 & -37.60 && 55485.3580 & -44.60 && 55881.3830 & -37.60 \\
55093.3480 & -37.50 && 55496.2960 & -42.40 && 55969.2210 & -39.80 \\
\tableline
&&& Hermes &&&& \\
\tableline
55023.6954 & -39.91 && 55767.6820 & -41.17 && 55872.4665 & -37.58 \\
55043.6739 & -41.23 && 55791.6949 & -39.06 && 55873.4721 & -37.46 \\
55045.6578 & -41.63 && 55828.5818 & -42.37 && 55881.3580 & -38.21 \\
55082.5616 & -37.26 && 55841.4558 & -43.04 && 55882.3858 & -38.59 \\
55106.5933 & -42.93 && 55842.4490 & -43.12 && 55883.4060 & -38.81 \\
55168.4603 & -40.99 && 55852.4188 & -42.09 && 55884.4160 & -39.02 \\
55394.7242 & -45.04 && 55862.4199 & -39.66 && 55886.3551 & -39.13 \\
55413.6809 & -43.12 && 55866.4695 & -38.99 && 55888.4476 & -38.99 \\
55429.5996 & -41.78 && 55867.4144 & -38.64 && 55889.4222 & -38.97 \\
55474.6315 & -45.73 && 55868.4160 & -38.57 && 55934.3511 & -44.37 \\
55502.4670 & -41.36 && 55870.4351 & -37.93 && 55937.3255 & -44.09 \\
55760.6214 & -42.20 && 55871.4217 & -37.70 && 55959.3281 & -41.13 \\
\enddata
\end{deluxetable}

\clearpage

\tablenum{6}
\begin{deluxetable}{cccccccc}
\tablecaption{Radial Velocity Observations of
IRAS 22272+5435\label{tab_22272RV}}
\tablewidth{0pt} \tablehead{ \colhead{HJD$-$2,400,000}
&\colhead{{\it V}$_ r$} 
&&\colhead{HJD$-$2,400,000} &\colhead{{\it V}$_ r$} 
&&\colhead{HJD$-$2,400,000}
&\colhead{{\it V}$_ r$} \\
\colhead{}&\colhead{(km s$^{-1}$)}&&\colhead{}&\colhead{(km s$^{-1}$)}&&\colhead{}&\colhead{(km s$^{-1}$)}  }
\startdata
&&& DAO-RVS &&&& \\
\tableline
47445.8630\tablenotemark{a} & -39.65 && 48627.6698 & -39.49 && 49177.9472 & -33.91 \\
47763.9090\tablenotemark{a} & -37.42 && 48719.9755 & -40.06 && 49189.8065 & -36.55 \\
47789.7860\tablenotemark{a} & -35.13 && 48733.9496 & -39.76 && 49217.9122 & -37.57 \\
48089.8730\tablenotemark{a} & -41.98 && 48756.9420 & -37.56 && 49239.9008 & -39.47 \\
48467.9460\tablenotemark{a} & -40.55 && 48779.9443 & -38.93 && 49263.9136 & -40.48 \\
48469.9154 & -39.46 && 48788.9657 & -37.96 && 49286.9238 & -35.72 \\
48470.9584 & -38.51 && 48799.9447 & -39.32 && 49554.9153 & -38.48 \\
48471.9298 & -39.04 && 48837.8984 & -38.67 && 49555.9754 & -38.49 \\
48510.8982 & -39.09 && 48875.8749 & -38.78 && 49899.9641 & -35.84 \\
48532.8379 & -36.89 && 48886.7604 & -36.41 && 49913.9708 & -36.21 \\
48533.8327 & -36.50 && 48922.8231 & -33.79 && 49940.8052 & -36.12 \\
48566.7707 & -33.79 && 49130.9400 & -36.10 && 50020.8213 & -38.46 \\
48588.7761 & -37.08 && 49163.9505 & -33.76 && \nodata & \nodata \\
\tableline
&&& DAO-CCD &&&& \\
\tableline
54327.9436 & -38.22 && 54796.6802 & -39.16 && 55406.9180 & -39.54 \\
54338.8345 & -40.66 && 54945.9563 & -41.91 && 55426.9012 & -38.16 \\
54354.9248 & -42.43 && 55014.8608 & -43.75 && 55432.8985 & -37.34 \\
54384.7966 & -42.12 && 55028.8458 & -43.48 && 55476.7658 & -42.91 \\
54391.7832 & -44.26 && 55041.8470 & -42.30 && 55482.8333 & -41.69 \\
54434.7340 & -38.41 && 55050.9506 & -42.12 && 55524.6949 & -42.22 \\
54658.9572 & -41.89 && 55055.9536 & -42.81 && 55719.8822 & -40.89 \\
54690.9112 & -42.31 && 55063.9118 & -42.93 && 55812.8778 & -42.59 \\
54709.9258 & -40.27 && 55076.9132 & -41.76 && 55880.7320 & -40.87 \\
54719.9920 & -39.04 && 55096.8600 & -39.90 && 55881.6898 & -40.92 \\
54725.9834 & -38.84 && 55167.7288 & -41.66 && 55897.7456 & -38.59 \\
54741.0134 & -38.95 && 55309.9507 & -37.01 && 55908.7460 & -38.50 \\
54749.8461 & -38.98 && 55322.9346 & -37.36 && 55938.6723 & -40.21 \\
54754.8387 & -38.49 && 55385.8899 & -38.99 && 55972.6684 & -40.16 \\
54790.7215 & -39.00 && 55392.8860 & -40.14 && \nodata & \nodata \\
\tableline
&&& CORAVEL &&&& \\
\tableline
54747.3470 & -36.50 && 55220.1920 & -37.90 && 55712.5220 & -38.00 \\
54748.2940 & -36.90 && 55221.1890 & -37.90 && 55719.5100 & -39.80 \\
54749.4310 & -37.40 && 55302.5870 & -34.60 && 55755.4500 & -38.10 \\
54764.4330 & -39.20 && 55346.5250 & -41.50 && 55772.5170 & -37.50 \\
54814.3590 & -38.50 && 55365.5080 & -40.00 && 55777.4960 & -39.20 \\
54815.2630 & -37.90 && 55394.5070 & -40.30 && 55778.5300 & -39.30 \\
54864.2080 & -37.80 && 55395.5210 & -39.80 && 55800.5130 & -42.10 \\
54866.2050 & -38.20 && 55399.5290 & -38.00 && 55801.4490 & -43.00 \\
54929.5980 & -41.80 && 55400.5330 & -39.20 && 55808.4930 & -39.80 \\
54940.5850 & -41.30 && 55435.4770 & -34.70 && 55815.4610 & -39.30 \\
54942.5720 & -40.90 && 55444.4970 & -35.40 && 55816.4010 & -39.00 \\
54947.5580 & -39.60 && 55446.5430 & -36.70 && 55826.4060 & -38.40 \\
54950.5640 & -40.40 && 55448.5490 & -36.60 && 55830.3660 & -40.90 \\
55000.4940 & -44.40 && 55463.4270 & -39.50 && 55852.4330 & -42.00 \\
55013.4750 & -42.10 && 55470.4590 & -41.40 && 55853.3390 & -43.90 \\
55027.5130 & -41.20 && 55473.5210 & -40.10 && 55860.3300 & -43.10 \\
55030.4820 & -41.80 && 55476.4210 & -41.50 && 55861.4080 & -43.10 \\
55037.5350 & -39.30 && 55477.3320 & -41.20 && 55875.3750 & -41.40 \\
55069.4380 & -41.10 && 55482.4210 & -40.30 && 55881.3930 & -39.40 \\
55076.4360 & -40.30 && 55485.3650 & -40.90 && 55969.2070 & -38.70 \\
55077.3690 & -40.70 && 55496.3150 & -39.40 && 55974.2210 & -38.20 \\
55092.4930 & -37.90 && 55532.3190 & -40.60 && 55994.2390 & -41.00 \\
55093.4240 & -36.90 && 55671.5830 & -40.00 && 55995.2310 & -40.60 \\
55095.4180 & -37.30 && 55675.5740 & -38.60 && 56029.5940 & -36.00 \\
55120.3500 & -38.00 && 55693.5480 & -37.20 && 56035.5860 & -35.00 \\
55124.3210 & -38.60 && 55696.5380 & -36.70 && \nodata & \nodata \\
55138.2570 & -40.90 && 55707.5270 & -36.80 && \nodata & \nodata \\
\tableline
&&& Hermes &&&& \\
\tableline
55003.6518 & -43.10 && 55760.6111 & -38.43 && 55873.4816 & -42.55 \\
55031.6428 & -41.20 && 55767.6247 & -38.18 && 55881.5364 & -41.09 \\
55051.6155 & -42.93 && 55791.7112 & -44.83 && 55882.3729 & -40.83 \\
55083.5700 & -39.50 && 55828.5719 & -38.90 && 55883.3813 & -40.83 \\
55106.6049 & -35.68 && 55841.4668 & -41.16 && 55884.4060 & -40.62 \\
55125.4671 & -40.33 && 55842.4580 & -41.37 && 55886.3125 & -40.43 \\
55168.4443 & -40.47 && 55851.4109 & -44.69 && 55887.4003 & -40.28 \\
55217.3265 & -38.49 && 55862.4104 & -44.73 && 55888.4352 & -40.05 \\
55347.7349 & -39.69 && 55866.4789 & -44.08 && 55889.3846 & -39.87 \\
55395.6933 & -40.21 && 55867.4048 & -43.74 && 55903.3333 & -38.31 \\
55421.6207 & -36.97 && 55868.4066 & -43.64 && 55932.3418 & -37.97 \\
55470.5451 & -42.58 && 55870.4257 & -43.18 && 55937.3359 & -38.45 \\
55474.6014 & -42.45 && 55871.4120 & -42.95 && 55959.3548 & -39.28 \\
55502.4764 & -40.27 && 55872.4518 & -42.66 && \nodata & \nodata \\
\enddata
\tablenotetext{a}{Observations made by R.D. McClure with using a K-star mask.}
\end{deluxetable}

\clearpage

\tablenum{7}
\begin{deluxetable}{ccrrrrrrrrrrrrrr}
\tablecolumns{17} \tabletypesize{\scriptsize}
\tablecaption{Periodogram Study of the Light and Color Curves of IRAS 22272+5435\tablenotemark{a}\label{22272_lc_per}}
\rotate
\tabletypesize{\footnotesize} 
\tablewidth{0pt} \tablehead{ 
\colhead{Data} &\colhead{Filter} & \colhead{Years}&\colhead{No.} & \colhead{P$_1$}&\colhead{A$_1$} &\colhead{$\phi$$_1$\tablenotemark{b}}&\colhead{P$_2$}&
\colhead{A$_2$} &\colhead{$\phi$$_2$\tablenotemark{b}}&\colhead{P$_3$} &\colhead{A$_3$}&\colhead{$\phi$$_3$\tablenotemark{b}}&\colhead{P$_4$} &\colhead{A$_4$}&\colhead{$\phi$$_4$\tablenotemark{b}}\\
\colhead{Set} & & &\colhead{Obs.} & \colhead{(days)}&\colhead{(mag)} & &\colhead{(days)}&
\colhead{(mag)} & &\colhead{(days)} &\colhead{(mag)}& &\colhead{(days)} &\colhead{(mag)} 
& } \startdata
A& {\it V} & 1991-2011 & 507  &131.9 & 0.097 & 0.48 & 125.0& 0.048 & 0.94 & 145.4 & 0.044  & 0.48 & 229.5  & 0.042 & 0.97\\
B & {\it V} & 1991-1999 & 230 &131.4 & 0.110 & 0.42& 146.7& 0.055& 0.66 & 225.0 & 0.042 & 0.84 & 82.9  &0.035 & 0.20\\
C & {\it V} & 2002-2011 & 277  &131.4 & 0.102 & 0.28 & 125.5 & 0.072 & 0.16 & 152.1 & 0.056 & 0.75 & 248.8 & 0.043 & 0.50\\
D & {\it R}$_C$ & 2002-2011& 287 &131.2 & 0.076 & 0.16 & 125.8 & 0.058 & 0.23 & 164.4 & 0.045 & 0.32 & 138.1 & 0.039 & 0.78\\
E & {\it R}$_C$ & 1999-2011 & 335 &130.7 & 0.064 & 0.97 & 126.0 & 0.048 & 0.32 & 164.7 & 0.040 & 0.41 & 138.7& 0.037 & 0.01\\
F & {\it V$-$R}$_C$ & 2002-2011 & 265 &130.3 & 0.023 & 0.88& 124.2& 0.013& 0.65 & 138.1 & 0.012 & 0.69 & \nodata & \nodata & \nodata\\
\enddata
\tablenotetext{a}{The uncertainties in {\it P}, {\it A}, $\phi$ are approximately $\pm$0.3 day, $\pm$0.005 mag, $\pm$0.015, respectively, 
except for data set A, in which the uncertainty in {\it P} is $\pm$0.1 day, and set F, in which the uncertainty in A is $\pm$0.002.}
\tablenotetext{b}{The phases are determined based on the epoch of 2,448,000.00.}
\end{deluxetable}

%\clearpage

\tablenum{8}
\begin{deluxetable}{crrcrcrrcrrcr} 
\tablecolumns{13} \tabletypesize{\scriptsize}
\tablecaption{Periodogram Study of the Radial Velocity Curve of IRAS 22272+5435\tablenotemark{a}\label{22272_vc_per}}
\rotate
\tabletypesize{\footnotesize} 
\tablewidth{0pt} \tablehead{ 
\colhead{Data} & \colhead{Years}&\colhead{No.} & \colhead{A$_0$}& \colhead{P$_1$}&\colhead{A$_1$} &\colhead{$\phi$$_1$\tablenotemark{b}}&\colhead{P$_2$}& \colhead{A$_2$} &\colhead{$\phi$$_2$\tablenotemark{b}}&\colhead{P$_3$}& \colhead{A$_3$} &\colhead{$\phi$$_3$\tablenotemark{b}}\\
\colhead{Set} &\colhead{}&\colhead{Obs.} &\colhead{(km~s$^{-1}$)}& \colhead{(days)}&\colhead{(km~s$^{-1}$)} & &\colhead{(days)}& \colhead{(km~s$^{-1}$)} & &\colhead{(days)}& \colhead{(km~s$^{-1}$)} & } 
\startdata
A & 1988-2011 & 274 & $-$40.77\tablenotemark{c}&131.2 & 1.52 & 0.96 & 125.5 & 1.42 & 0.93 & 66.8 & 1.01 & 0.98 \\
B & 2005-2011 & 236 & $-$40.81 &132.7 & 1.43 & 0.54 & 67.0 & 1.13 & 0.36 & 125.8 & 1.04 & 0.03 \\
C & 1988-1995& 38 & $-$40.59\tablenotemark{c} &126.2 & 1.83 & 0.95 & 148.5 & 1.43 &0.35 & \nodata & \nodata & \nodata \\
\enddata
\tablenotetext{a}{The uncertainties in {\it P}, {\it A}, $\phi$ for Set A are approximately $\pm$0.1 day, $\pm$0.15 km~s$^{-1}$, $\pm$0.018, respectively.  They are larger for the other, smaller data sets.}
\tablenotetext{c}{The velocities from the 1988$-$1995 data set were adjusted with an additive offset of $-$2.80 km~s$^{-1}$ to bring them to approximately the same level as the zero-point adjusted 2005$-$2011 data.}
\tablenotetext{b}{The phases are determined based on the epoch of 2,448,000.00.}
\end{deluxetable}

\clearpage

\tablenum{9}
\begin{deluxetable}{ccrrrrrrrrrrrrrr}
\tablecolumns{17} \tabletypesize{\scriptsize}
\tablecaption{Periodogram Study of the Light and Color Curves of IRAS 22223+4327\tablenotemark{a}\label{22223_lc_per}}
\rotate
\tabletypesize{\footnotesize} 
\tablewidth{0pt} \tablehead{ 
\colhead{Data} &\colhead{Filter} & \colhead{Years}&\colhead{No.} & \colhead{P$_1$}&\colhead{A$_1$} &\colhead{$\phi$$_1$\tablenotemark{b}}&\colhead{P$_2$}&
\colhead{A$_2$} &\colhead{$\phi$$_2$\tablenotemark{b}}&\colhead{P$_3$} &\colhead{A$_3$}&\colhead{$\phi$$_3$\tablenotemark{b}}&\colhead{P$_4$} &\colhead{A$_4$}&\colhead{$\phi$$_4$\tablenotemark{b}}\\
\colhead{Set} & & &\colhead{Obs.} & \colhead{(days)}&\colhead{(mag)} & &\colhead{(days)}&
\colhead{(mag)} & &\colhead{(days)} &\colhead{(mag)}& &\colhead{(days)} &\colhead{(mag)} 
& } \startdata
A& {\it V} & 1994-2011 & 630  & 90.5 & 0.040 & 0.24 & 85.8 & 0.029 & 0.89 & 92.4 & 0.020  & 0.22 & 83.8  & 0.016 & 0.55\\
B & {\it V} & 1994-2002 & 210  & 89.7 & 0.052 & 0.00 & 83.3 & 0.027 & 0.17 & 96.2 & 0.021 & 0.06 & 81.0  & 0.018 & 0.86\\
C & {\it V} & 2003-2011 & 420  & 86.7 & 0.033 & 0.61 & 89.5 & 0.028 & 0.65 & 94.3 & 0.014 & 0.56 & \nodata & \nodata & \nodata\\
D & {\it R}$_C$ & 2003-2011 & 289  & 86.6 & 0.022 & 0.58 & 89.9 & 0.019 & 0.80 & \nodata & \nodata & \nodata & \nodata & \nodata & \nodata\\
E & {\it R}$_C$ & 1995-2011 & 387  & 90.5 & 0.026 & 0.24 & 85.7 & 0.018 & 0.86 & 67.8 & 0.014 & 0.90 & 74.3 & 0.012 & 0.49\\
F & {\it V$-$R}$_C$ & 1995-2011 & 332  &90.4 & 0.010 & 0.16 & 86.1 & 0.007 & 0.04 & 91.9 & 0.007 & 0.89 & 78.8 & 0.004 & 0.93\\
\enddata
\tablenotetext{a}{The uncertainties in {\it P} are approximately $\pm$0.07 day in data sets A, E, and F and  approximately $\pm$0.16 day in sets B, C, and D.  
The uncertainties in {\it A} and $\phi$ are approximately $\pm$0.002 mag and $\pm$0.015, respectively, 
except for set F, in which the uncertainty in {\it A} is $\pm$0.001 mag, and set A, in which the uncertainty in $\phi$ is $\pm$0.010.}
\tablenotetext{b}{The phases are determined based on the epoch of 2,448,000.00.}
\end{deluxetable}

%\clearpage

\tablenum{10}
\begin{deluxetable}{rrrrccrrrr}
\tablecolumns{16} \tabletypesize{\scriptsize}
\tablecaption{Periodogram Study of the Radial Velocity Curve of IRAS 22223+4327\tablenotemark{a}\label{22223_vc_per}}
\rotate
\tabletypesize{\footnotesize} 
\tablewidth{0pt} \tablehead{ 
\colhead{Data} & \colhead{Years}&\colhead{No.} & \colhead{A$_0$}& \colhead{P$_1$}&\colhead{A$_1$} &\colhead{$\phi$$_1$\tablenotemark{b}}&\colhead{P$_2$}&
\colhead{A$_2$} &\colhead{$\phi$$_2$\tablenotemark{b}}\\
\colhead{Set} &\colhead{}&\colhead{Obs.} &\colhead{(km~s$^{-1}$)}& \colhead{(days)}&\colhead{(km~s$^{-1}$)} & &\colhead{(days)}&
\colhead{(mag)} & } 
\startdata
A& 1991$-$2011 & 189 & $-$41.10 & 88.8 & 1.95 & 0.49 & \nodata & \nodata & \nodata \\
B & 2007$-$2011 & 147 & $-$41.20 & 88.3 & 1.99 & 0.96 & \nodata & \nodata & \nodata \\
C & 1991$-$1995 & 34 & $-$40.58 & 88.8\tablenotemark{c} & 2.20 & 0.50 & \nodata & \nodata & \nodata \\
\enddata
\tablenotetext{a}{The uncertainties in {\it P} vary quite a bit depending upon the length of the observing interval, and are approximately $\pm$0.05 day (Set A), $\pm$0.3 day (Set B), and
$\pm$0.5 day (Set C).  The uncertainties in {\it A} and $\phi$ are much more similar among the sets, and are  approximately $\pm$0.20 km~s$^{-1}$ and $\pm$0.02, respectively. }
\tablenotetext{b}{The phases are determined based on the epoch of 2,448,000.00.}
\tablenotetext{c}{This period determination does not quite reach our accepted level of significance, but is included to show the agreement with the other values.}
\end{deluxetable}

\clearpage

%\clearpage

%\clearpage

%\clearpage

%\clearpage

%\clearpage

%\clearpage

%\clearpage
 
%\clearpage

%\clearpage

%\clearpage

%\clearpage

%\clearpage

%\clearpage

%\clearpage
 
\clearpage

\title{ERRATUM: ``STUDIES OF VARIABILITY IN PROTO-PLANETARY NEBULAE: II. LIGHT AND VELOCITY CURVE ANALYSES OF IRAS 22272+5435 and 22223+4327'' (2013, ApJ, 766, 116)}


At the time of our publication of this paper, we were not aware of the recent photometric study of IRAS 22272+5435 by \citet{ark09} and thus did not cite it in our study.  They combined new observations from 2000$-$2008 with their previous data and carried out a period analysis of their entire data set.  They also find two closely spaced periods that beat against each other to produce the modulated light curves.  Their values are close to but not identical to ours.  They find periods of 128 and 131 day, and a third weaker one of 125 day; from their 2000$-$2008 data alone they find a dominant period of 128 day. These can be compared with our results, based on a larger data set of our 1994$-$2011 and their 1991$-$1999 measurements combined, of a dominant period of 131 day and a secondary period of 125 day. We did not find a period of 128 day, even when we analyzed our 2002$-$2011 data alone.

They also make a comparison between the light curves and radial velocity curves, in their case using only the 2006 radial velocity curves of \citet{zacs09}.  Contrary to the earlier statement by \citet{hri00}, based on a small data set, and the conclusions of \citet{ark09}, the star is not brightest at average size and expanding, but is actually brightest when smallest and faintest when largest, as seen in our Figure 7 and also in their Figure 4.

\end{document}